\documentclass[a4paper, review, 10pt]{article}

\usepackage{lineno,hyperref}

\usepackage{latexsym}

\usepackage{url}
\usepackage{hyperref}

\usepackage[utf8]{inputenc}
\usepackage{amsmath}
\usepackage{amsfonts}
\usepackage{amssymb}
\usepackage{graphicx}
\usepackage{multirow}
\usepackage{subfigure}
\usepackage{bm}
\usepackage{array}

\usepackage{color}
\usepackage[figuresright]{rotating}
\usepackage{threeparttable}
 \usepackage{float}

\def\etal{{\em et al.}}
\def\eg{{\em e.g.}}
\def\ie{{\em i.e.}}
\def\etc{{\em etc.}}

\author{Li Zhang$^1$, Mingliang Wang$^2$, Mingxia Liu$^{3,*}$ and Daoqiang Zhang$^{2,*}$}

\title{A Survey on Deep Learning for Neuroimaging-based Brain Disorder Analysis}

\begin{document}

\maketitle

{\small \begin{enumerate}
	\item College of Computer Science and Technology, Nanjing Forestry University, Nanjing 210037, China.
	\item  College of Computer Science and Technology, Nanjing University of Aeronautics and Astronautics, Nanjing 211106, China.
	\item Department of Radiology and BRIC, University of North Carolina at Chapel Hill, NC 27599, USA.
\end{enumerate}}

\begin{abstract}
Deep learning has been recently used for the analysis of neuroimages, such as structural magnetic resonance imaging (MRI), functional MRI, and positron emission tomography (PET), and has achieved significant performance improvements over traditional machine learning in computer-aided diagnosis of brain disorders.
This paper reviews the applications of deep learning methods for neuroimaging-based brain disorder analysis. We first provide a comprehensive overview of deep learning techniques and popular network architectures, by introducing various types of deep neural networks and recent developments.
We then review deep learning methods for computer-aided analysis of four typical brain disorders, including Alzheimer's disease, Parkinson's disease, Autism spectrum disorder, and Schizophrenia, where the first two diseases are neurodegenerative disorders and the last two are neurodevelopmental and psychiatric disorders, respectively. More importantly, we discuss the limitations of existing studies and present possible future directions.

\end{abstract}


\section{Introduction}

Medical imaging refers to several different technologies that are used to provide visual representations of the interior of the human body in order to aid the radiologists and clinicians to early detect, diagnose or treat diseases more efficiently~\cite{brody2013medical}. Over the past few decades, medical imaging quickly becomes a dominant and effective tool, and represents various imaging modalities, including X-ray, mammography, ultrasound, computed tomography, magnetic resonance imaging(MRI) and positron emission tomography(PET)~\cite{heidenreich2002modern}. Each type of these technologies gives various anatomical and functional information about the different body organs for diagnosis as well as for research. In clinical practice, the detail interpretation of medical images needs to be performed by human experts such as the radiologists and clinicians. However, for the enormous number of medical images, the interpretations are time-consuming and easily cause by the biases and potential fatigue of human experts. Therefore, from the early 1980s, doctors and researchers have begun to use computer-assisted diagnosis(CAD) systems to interpret the medical images and to improve their efficiency.

In the CAD systems, machine learning is able to extract informative features which well describe the inherent patterns from data, and plays a vital role in medical image analysis~\cite{wernick2010machine, erickson2017machine}. Several traditional machine learning algorithms, such as sparse learning, support vector machine (SVM), Gaussian networks, random forest, decision tree and hidden Markov model, \etc, are wildly used~\cite{wu2016machine}.
However, the structures of the medical images are very complex, and the feature selection step is still done by the human experts on the basis of their domain-specific knowledge. This results in a challenge for non-experts to utilize machine learning techniques in medical image analysis. Therefore, the handcrafted feature selection is not suitable for medical images. In addition, the shallow architectures of these traditional machine learning algorithms limit their representational power~\cite{pandya2019medical}.

Naturally, the logical next research direction is to let algorithms automatically learn features that can optimally represent the data.
 Deep learning perfectly coincides with this concept and rapidly becomes a methodology of choice for medical image analysis in recent years~\cite{goodfellow2016deep,lecun2015deep,schmidhuber2015deep}.
Compared to the traditional machine learning algorithms, deep learning automatically discoveries the informative representations without the professional knowledge of domain experts and allows the non-experts to effectively use deep learning techniques.
Due to enhanced computer power with the high-tech central processing units(CPU) and graphical processing units(GPU), given the availability of big data, and designed novel algorithms to train deep neural networks, deep learning receives the unprecedented success in the most artificial intelligence applications, such as computer vision~\cite{voulodimos2018deep}, natural language processing~\cite{sarikaya2014application} and speech recognition~\cite{bahdanau2016end}. Especially, the improvement and successes of computer vision simultaneously prompted the use of deep learning in the medical image analysis~\cite{lee2017deep,shen2017deep}.

Currently, deep learning has fueled great strides in medical image analysis. We can divide the medical image analysis tasks into several major categories: classification, detection/localization, registration, segmentation~\cite{litjens2017survey}.
The classification is one the first task in which deep learning giving a major contribution to medical image analysis. This task aims to classify medical images into two or more classes.
The stacked auto-encoder model was used to identify Alzheimer's disease or mild cognitive impairment by combining medical images and biological features~\cite{suk2015latent}.
The detection/localization task consists of the localization and identification of the landmarks or lesion in the full medical image. 
For example, deep convolutional neural networks were used for the detection of lymph nodes in CT images~\cite{roth2014new}.
The segmentation task is to partition a medical image into different meaningful segments, such as different tissue classes, organs, pathologies, or other biologically relevant structures.
The U-net was the most well-known deep learning architecture, which used convolutional networks for biomedical image segmentation~\cite{ronneberger2015u}.
Registration of medical images is a process that searches for the correct alignment of images.
Wu~\etal~utilized convolutional layers to extract features from input patches in an unsupervised manner. Then the obtained feature vectors were used to replace the handcrafted features in the HAMMER registration algorithm~\cite{wu2013unsupervised}. In addition, the medical image analysis contains other meaningful tasks, such as content-based image retrieval~\cite{li2018large}, image generation and enhancement~\cite{oktay2016multi}, combination image data with reports~\cite{schlegl2015predicting}.

There are many papers have comprehensively surveyed the medical image analysis using deep learning techniques~\cite{lee2017deep,shen2017deep,litjens2017survey}. However, these papers usually reviewed all human tissues including the brain, chest, eye, breast, cardiac, abdomen, musculoskeletal and other. Almost no papers focus on one specific tissue or disease~\cite{hu2018deep}.
Brain disorders are among the most severe health problems facing our society, causing untold human suffering and enormous economic costs. Many studies successfully used medical imaging techniques for the early detection, diagnosis and treatment of the human brain disorders, such as neurodegenerative disorders, neurodevelopmental disorders and psychiatric disorders~\cite{vieira2017using, durstewitz2019deep}. Therefore, we pay more close attention to human brain disorders in this review.

The structure of this review can roughly be divided into two parts, the deep learning architectures and the usage of deep learning in brain disorder analysis, and is organized as follows. In Section~\ref{S2}, we briefly introduce some popular deep learning models. In Section~\ref{S3}, we provide a detailed overview of recent studies using deep learning techniques for four brain disorders, including Alzheimer's disease, Parkinson's disease, Autism spectrum disorder, and Schizophrenia. Finally, we analyze the limitations of the deep learning techniques in medical image analysis, and provide some research directions for further study.  For readers' convenience, the abbreviations of terminologies used in the following context are listed in the ~\ref{tab:Abbr}.

\begin{table}[!htbp]  
		\renewcommand\arraystretch{1.1}
	\scriptsize
	\caption{Abbreviation of terminologies used in following sections.}
	\centering
	\begin{tabular*}{0.98\textwidth}{@{\extracolsep{\fill}}ll||ll}
		\hline
		Terminology & Abbr. &	Terminology & Abbr. \\\hline
		Alzheimer's Disease  & AD     &  	Two-dimensional CNN  & 2D-CNN  \\
		Autism spectrum disorder & ASD  &	Three-dimensional CNN & 3D-CNN \\		
		Computer-assisted diagnosis & CAD &	Auto-encoder & AE       \\
		Converted  MCI  & cMCI &		Artificial neural networks & ANN    \\
		Cerebrospinal fluid &  CSF  &          Back-propagation & BP   \\
		Computed tomography&  CT  &	Convolutional neural networks  & CNN \\	
		Diffusion tensor imaging &  DTI & Denoising auto-encoders   &  DAE \\		
		Electroencephalogram   & EEG   & Deep belief networks &  DBN  \\	
		Functional MRI  & fMRI   &	Deep bolztman machine &  DBM \\		   	
		Gray matter & GM     &				Deep generative model & DGM  \\	
		Mild cognitive impairment & MCI  &Deep neural networks  & DNN \\		
		Magnetic resonance imaging & MRI  &	Feed-forward neural network  &  FFNN \\
		Normal control & NC   &   Generative adversarial networks  &  GAN \\	
		Parkinson's  Disease  &  PD  &		Graph CNN  & GCN \\			
		Positron emission tomography & PET    & 	Gated recurrent unit &  GRU  \\		 
		Region of interest   & ROI   &		Long-short term memory & LSTM    \\		
		Resting-state fMRI  & rs-fMRI  &	Multi-layer perceptron & MLP     \\		
		Stable MCI  & sMCI   & Principle component analysis & PCA  \\			
		Structural MRI  &  sMRI  &	Restricted Bolztman machine & RBM \\		
		Single photon emission CT  & SPECT  &	    Recurrent neural networks & RNN \\			
		Schizophrenia  & SZ &	 			Region with CNN  & R-CNN  \\	 
		White matter & WM &		 	Stacked auto-encoders   & SAE   \\
		&&  Stacked sparse AE & SSAE\\
		&&  	Support vector machine & SVM    \\
		&&  	Variational auto-encoders  &VAE      \\
		\hline 
	\end{tabular*}
	\begin{flushleft}
		The abbreviations in the  left column relate to the medical image analysis, and the abbreviations in the right column relate to the machine learning and deep learning architectures.
		
	\end{flushleft}
	
	\label{tab:Abbr}
\end{table}

\section{Deep Learning}
\label{S2}
In this section, we introduce the fundamental concept of basic deep learning models in the literature, which have been wildly applied to medical image analysis, especially human brain disorder diagnosis. These models include feed-forward neural networks, stacked auto-encoders, deep belief network, deep Boltzmann machine, generative adversarial networks, convolutional neural networks, graph convolutional networks and recurrent neural networks.

\subsection{Feed-Forward Neural Networks}
\begin{figure}[h]
	\centering
	\subfigure[]{\includegraphics[height=0.3\textwidth,angle=0]{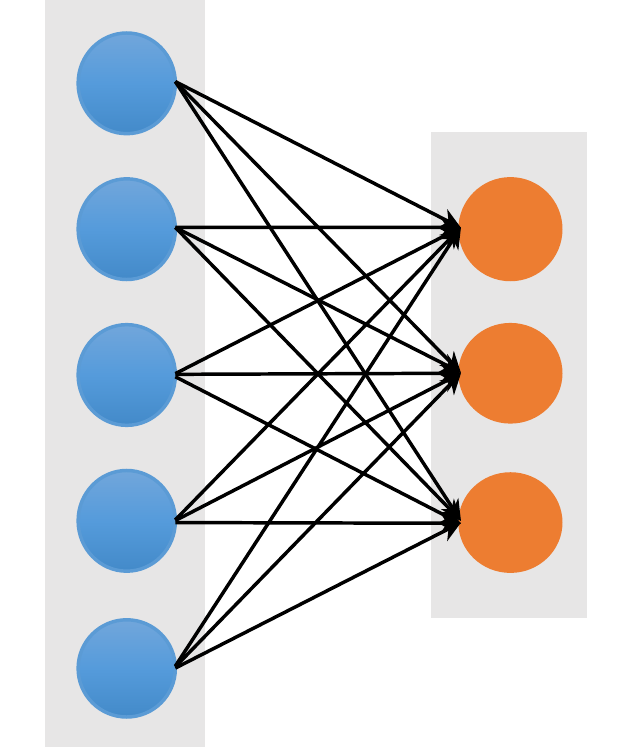}}
	\hfil
	\subfigure[]{\includegraphics[height=0.3\textwidth,angle=0]{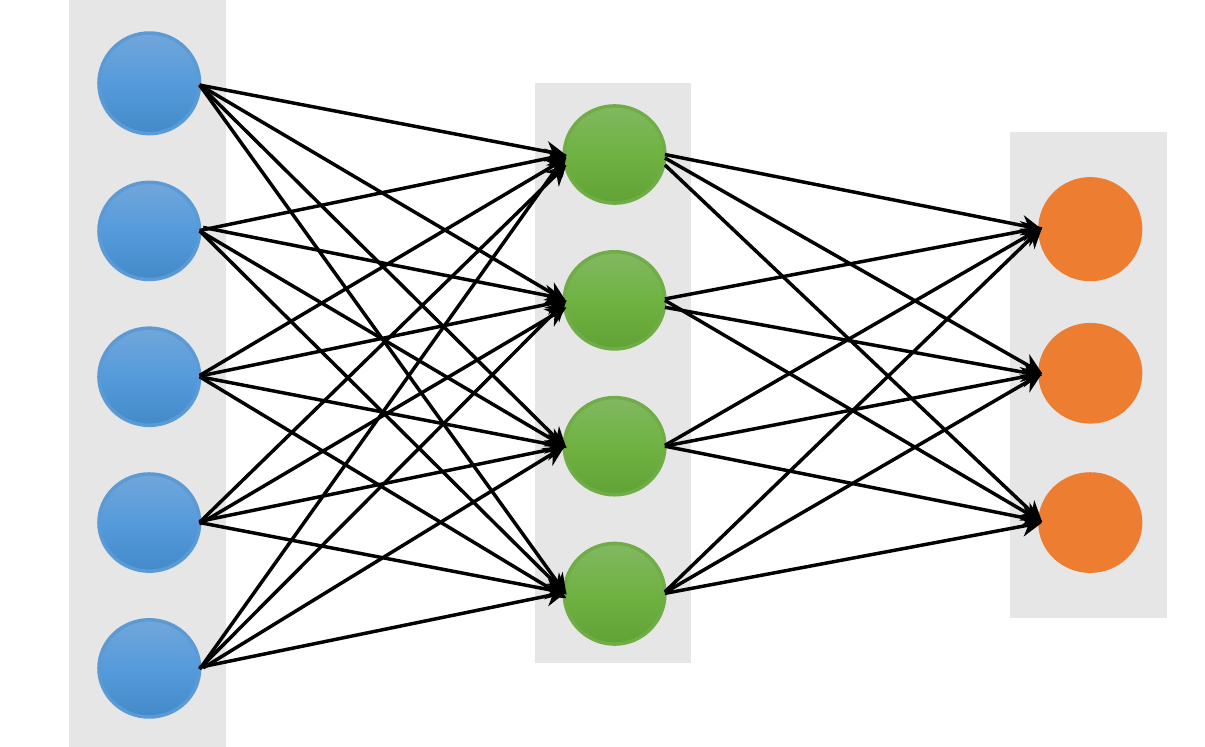}}
	\caption{Architectures of the single-layer(a) and multi-layer(b) neural networks. The blue, green and orange solid circles represent the input visible, hidden and output units respectively.
	}
	\label{fig-FFNN}
\end{figure}

In machine learning, artificial neural networks (ANN) aims to simulate intelligent behavior by mimicking the way that biological neural networks function.
The simplest artificial neural networks is a single-layer architecture, which is composed of an input layer and an output layer (Figure~\ref{fig-FFNN}.a).
However,  despite the use of non-linear activation functions in output layers, the single-layer neural network usually obtains poor performance for complicated data patterns.
In order to circumvent the limitation, the multi-layer perceptron (MLP), also referred to as a feed-forward neural network (FFNN) (Figure~\ref{fig-FFNN}.b), which includes a so-call hidden layer between the input layer and the output layer.  Each layer contains multiple units which are fully connected to units of neighboring layers, but there are no connections between units in the same layer. Given an input visible vector $\bm{x}$, the composition function of output unit $\bm{y}$ can be written as follows:
\begin{equation}\label{sq:mlp1}
y_{k}(\bm{x};\bm{\theta})=f^{(2)}\left(\sum_{j=1}^{M}w_{k,j}^{(2)}f^{(1)}\left( \sum_{i=1}^{N}w_{j,i}^{(1)}x_{i}+b_{j}^{(1)}\right)+b_{k}^{(2)}\right)
\end{equation}
where the superscript represents a layer index, $M$ is the number of hidden units, $b_{j}$ and $b_k$ represents the bias of input and hidden layer respectively.  $f^{(1)}(\cdot)$ and $f^{(2)}(\cdot)$ denote the nonlinear activation function and  the parameter set  is $\bm{\theta}=\{\bm{w}_{j}^{(1)},\bm{w}_{k}^{(2)},b_{j}^{(1)},b_{k}^{(2)}\}$.  The  back-propagation(BP) is an efficient algorithm to evaluate a gradient in the FFNN~\cite{rumelhart1986learning}.
The BP algorithm is to propagate the error values from the output layer back to the input layer through the network.
Once the gradient vector of all the layers is obtained, the parameters $\bm{\theta}$ can be updated. Until the loss function is converged or the predefined number of iterations is reached, the update process stops and the network gets the model parameters $\bm{\theta}$.

\subsection{Stacked Auto-Encoders}
\begin{figure}[h]
	\centering
	\includegraphics[scale=0.4]{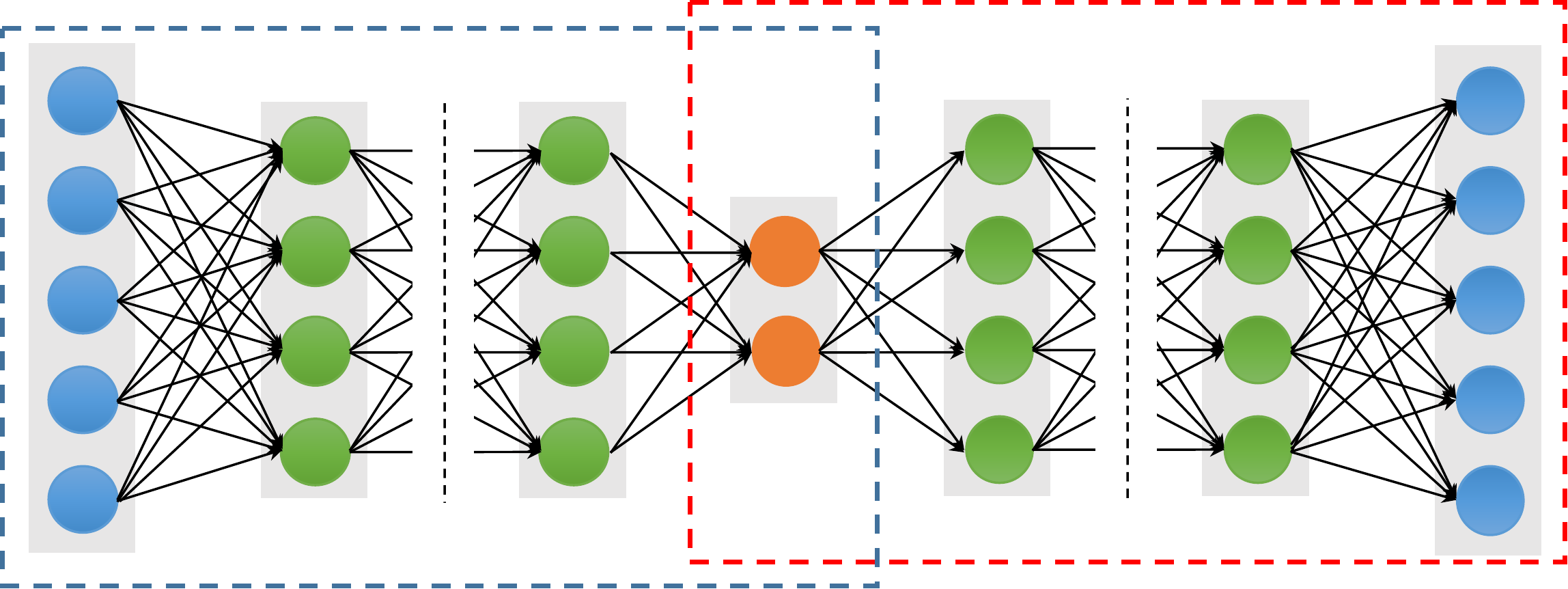}
	\caption{Architectures of  a stacked auto-encoder. The blue and red dotted boxes represent the encoding an decoding stage respectively. }
	\label{Fig-AE}
\end{figure}

An auto-encoder(AE), also known as auto-associator, learns the latent representations of input data (called encode) in an unsupervised manner, and then uses these representations to reconstruct output data (called decode).
Due to the simple and shallow structure, the power representation of a typical AE is relatively limited.
However, when multiple AEs are stacked to form a deep network, called a stacked auto-encoders (SAE)~(Figure~\ref{Fig-AE}), the representation power of an SAE can be obviously improved~\cite{hinton2006reducing}.
Because of the deep structural characteristic, the SAE is able to learn and discover more complicated patterns inherent in the input data. The lower layers can only learn simpler data patterns, while the higher layers are able to extract more complicated data patterns. In a word, the different layers of an SAE represent different levels of data information.
In addition, various AE variations, denoising auto-encoders (DAE)~\cite{vincent2008extracting},  sparse auto-encoder (sparse AE)~\cite{poultney2007efficient} and variational auto-encoders(VAE)~\cite{kingma2013auto}, have been proposed and also can be stacked as SAE, such as the stacked sparse AE (SSAE)~\cite{shin2013stacked} . These extensions of auto-encoders not only can learn more useful latent representations but also improve the robustness.

With respect to training parameters of an SAE, the greedy layer-wise approach is a well-known choice to learn the weight matrices and the biases.
It can avoid the drawback of the BP algorithm, which can cause the gradient falling into a poor local optimum~\cite{larochelle2009exploring}. The important character of the greedy layer-wise is to pre-train each layer in turn.
In other words, the output of the $l$-th hidden layers is used as input data for the $(l+1)$-th hidden layer. The process is performed as pre-training, which is conducted in an unsupervised manner with a standard BP algorithm. The important advantage of the pre-training is able to increase the size of the training dataset using unlabeled samples.

\subsection{Deep Belief Networks}
\begin{figure*}[h]
	\centering
	\subfigure[Deep Belief Network]{\includegraphics[width=0.35\textwidth]{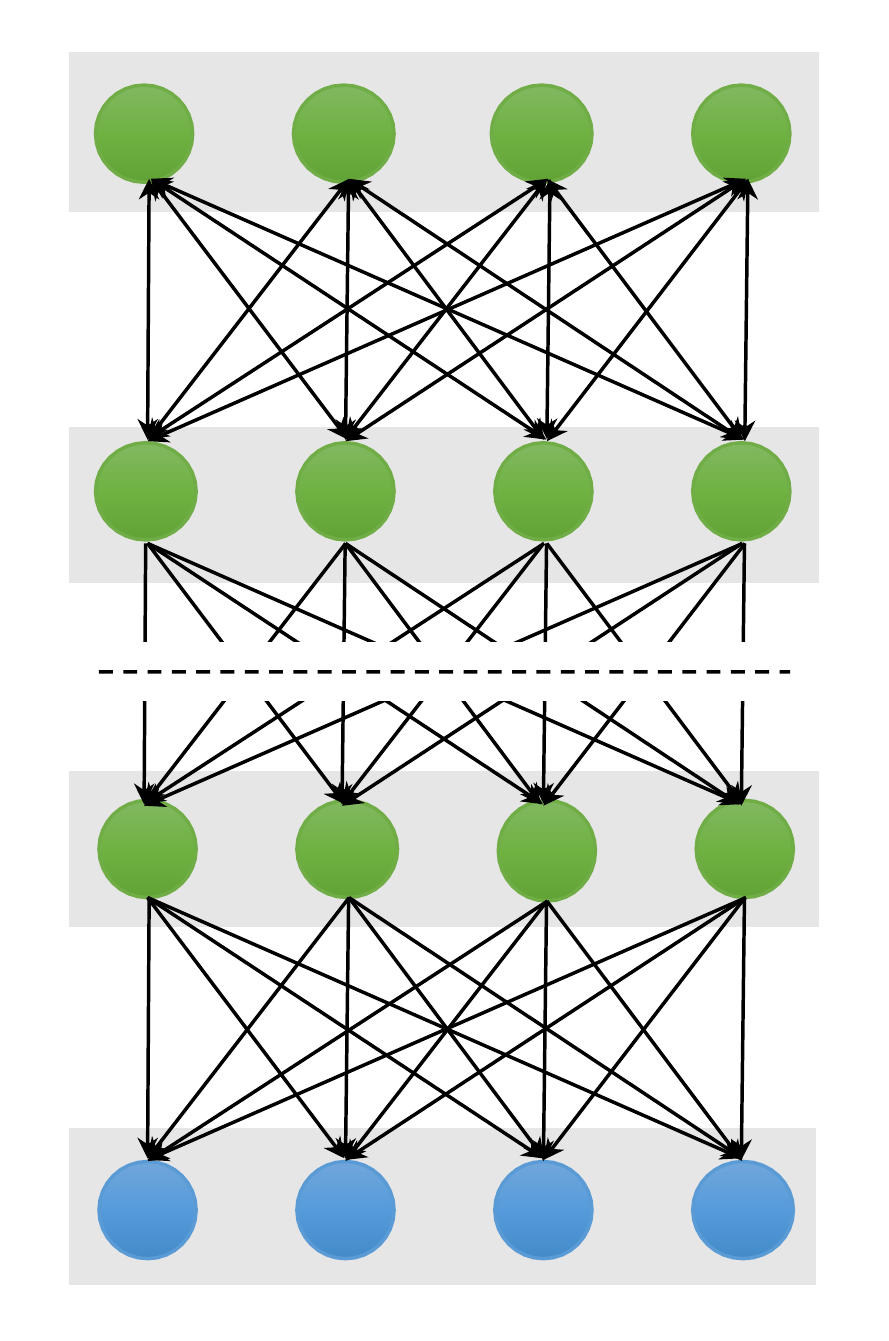}}
	\hfil
	\subfigure[Deep Boltzman Machine]{\includegraphics[width=0.35\textwidth]{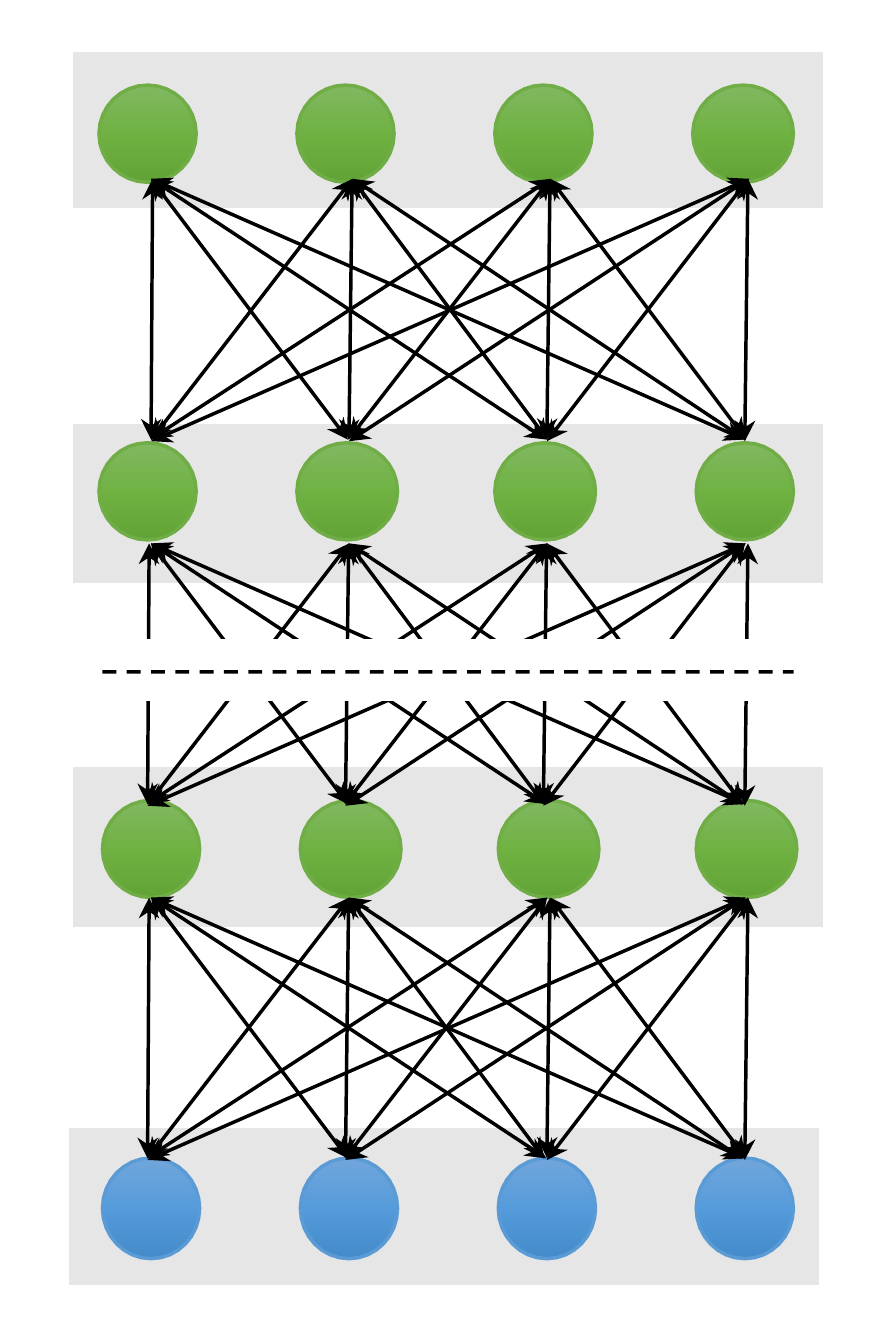}}
	\caption{Schematic illustration of Deep Belief Networks and Deep Boltzmann Machine
	}
	\label{Fig-BN}
\end{figure*}

A Deep Belief Networks (DBN), stacks multiple RBMs for deep architecture construction~\cite{hinton2006fast}.
A DBN has one visible layer and multiple hidden layers as shown in Figure~\ref{Fig-BN}.a.
The lower layers form directed generative models. However, the top two layers form the distribution of RBM, which is an undirected generative model.
Therefore, given the visible units $\bm{v}$ and $L$ hidden layers $\bm{h}^{(1)},\bm{h}^{(2)},\dots,\bm{h}^{(L)}$, the joint distribution of DBN is defined as:
\begin{equation}
P(\bm{v},\bm{h^{(1)}},\dots,\bm{h^{(L)}})=P(\bm{v}|\bm{h^{(1)}})\big(\prod_{l=1}^{L-2}P(\bm{h^{(l)}}|\bm{h^{(l+1)}})\big)P(\bm{h^{(L-1)}},\bm{h^{(L)}})
\end{equation}
where $P(\bm{h^{(l)}}|\bm{h^{(l+1)}})$ represents the conditional distribution for the units of the hidden layer $l$ given the units of the hidden layer $l+1$,
and $P(\bm{h^{(L-1)}}, \bm{h^{(L)}})$ corresponds the joint distribution of the top hidden layers $L-1$ and $L$.

As for training DBN, there are two steps, including pre-training and fine-tuning. In the pre-training step, DBN is trained by stacking RBMs layer by layer to find the parameter space. Each layer is trained as an RBM. Specifically, the $l$-th hidden layer is trained as RBM using the observation data from output representation of the $(l-1)$-th hidden layer, then repeats training each layer until the top layer. After the pre-training is completed, the fine-tuning is performed to further optimize the network to search the optimum parameters.
The wake-sleep algorithm and the standard back-propagation algorithm are good at fine-tuning for generative and discriminative models respectively~\cite{hinton1995wake}. For a practical application problem, the obtained parameters from the pre-training step are used to initiate a DNN, and then the deep model can be fine-tuned by a supervised learning algorithm like BP.

\subsection{Deep Boltzmann Machine}
A Deep Boltzmann Machine(DBM) is also constructed by stacking multiple RBMs as shown in Figure~\ref{Fig-BN}.b~\cite{salakhutdinov2010efficient, salakhutdinov2015learning}.
However, unlike DBN, all the layers of DBM form an entirely undirected model, and each variable within the hidden layers are mutually independent. Thus, the hidden layer $l$ is conditioned on its two neighboring layer $l-1$ and $l+1$, and its probability distribution is $P(\bm{h}^{(l)}|\bm{h}^{(l-1)},\bm{h}^{(l+1)} )$.
Given the values of the neighboring layers, the conditional probabilities over the visible and the $L$ set of hidden units are given by logistic sigmoid functions:
\begin{align}
P(v_i|\bm{h}^1) &= \sigma\big(\sum_{j} W_{ij}^{(1)}h_{j}^{(1)}\big)\\
P(h_{k}^{(l)}|\bm{h}^{(l-1)},\bm{h}^{(l+1)}) &= \sigma \big( \sum_{m} W_{mk}^{(l)} h_{m}^{(l-1)} + \sum_{n} W_{kn}^{(l+1)} h_{n}^{(l+1)} \big)\\
P(h_{t}^{(L)}|\bm{h}^{(L-1)}) &= \sigma \big( \sum_{s} W_{st}^{(L)} h_{s}^{(L-1)}  \big)
\end{align}
Note that in the computation of the conditional probability of the hidden unit $\bm{h}^{({l})}$, the probability incorporate both the lower hidden layer $\bm{h}^{({l-1})}$ and the upper hidden layer $\bm{h}^{(l+1)}$. Due to incorporate the more information from the lower and upper layers, the representational power of a DBM is more robust to the noisy observed data~\cite{karhunen2015unsupervised}. However, this character makes that the conditional probability of DBM $P(\bm{h}^{(l)}|\bm{h}^{(l-1)},\bm{h}^{(l+1)} )$ is more complex than those of DBN, $P(\bm{h}^{(l)}|\bm{h}^{(l+1)} )$.


\subsection{Generative Adversarial Networks }
\begin{figure}
	\centering
	\includegraphics[width=0.8\textwidth]{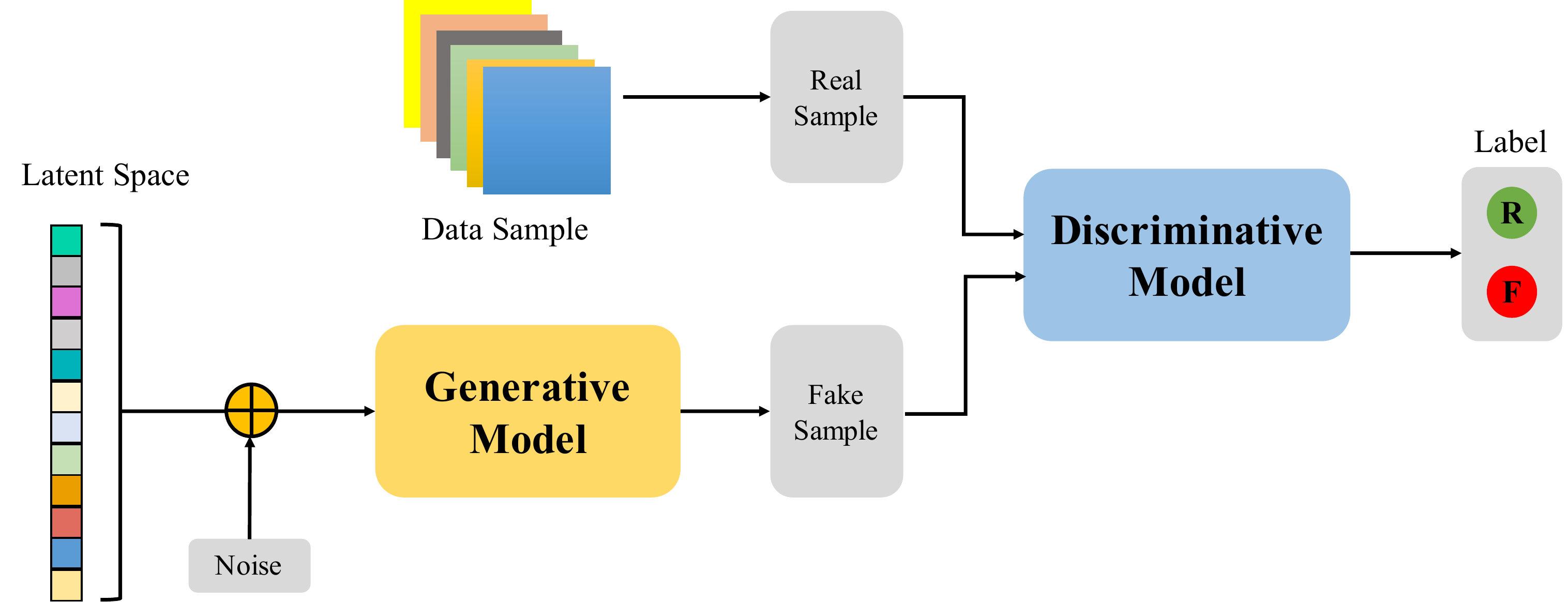}
	\caption{Architecture of Generative Adversarial Networks}
	\label{Fig-GAN}
\end{figure}

Due to their capability in learning deep representations without extensively annotated training data, Generative Adversarial Networks (GAN) has gained a lot of attention in computer vision and natural language processing~\cite{goodfellow2014generative}.
GAN consists of two competing neural networks, a generator $G$ and a discriminator $D$, as shown in Figure~\ref{Fig-GAN}. The generator $G$ parameterized by $\theta$ takes as input a random noise vector $\bm{z}$ from a prior distribution $p_{\bm{z}}(\bm{z};\theta)$, and outputs a sample $G(\bm{z})$, which can be regarded as a sample drawn from the generator data distribution $p_{g}$. The discriminator $D$ that takes an input $G(\bm{z})$ or $\bm{x}$, and outputs the probability $D(\bm{x})$ or $D(\bm{G(\bm{z})})$ to evaluate that the sample is from the generator $G$ or the real data distribution.
GAN simultaneously trains the generator and discriminator where the generator $G$ tries to generate realistic data to fool the discriminator, while the discriminator $D$ tries to distinguish between the real and fake samples. Inspired by the game theory, the training process is to form a two-player minimax game with the value function as follow:
\begin{equation}
\min_{G}\max_{D} V(G,D) = \mathbb{E}_{\bm{x} \sim p_{data}(\bm{x})} [\log D(\bm{x})] + \mathbb{E}_{\bm{z} \sim p_{\bm{z}}(\bm{z})}[\log(1-D(G(\bm{z})))]
\end{equation}
where $p_{data}(\bm{x})$ denotes the real data distribution. After training alternately, if $G$ and $D$ have enough capacity, they will reach a point at which both cannot improve because $p_{g}=p_{data}$. In other words, the discriminator is unable to distinguish the difference between a real and a generated sample, \ie $D(\bm{x})=0.5$.
Although vanilla GAN has attracted considerable attention in various applications, there still remain several challenges related to training and evaluating GAN, such as model collapse and saddle points~\cite{creswell2018generative}.
Therefore,  many variants of GAN, such as WGAN~\cite{arjovsky2017wasserstein} and DCGAN~\cite{radford2015unsupervised} have been proposed to overcome these challenges.

\subsection{Convolutional Neural Networks}
\begin{figure*}[htbp]
	\centering
	\includegraphics[width=0.85\textwidth,angle=0]{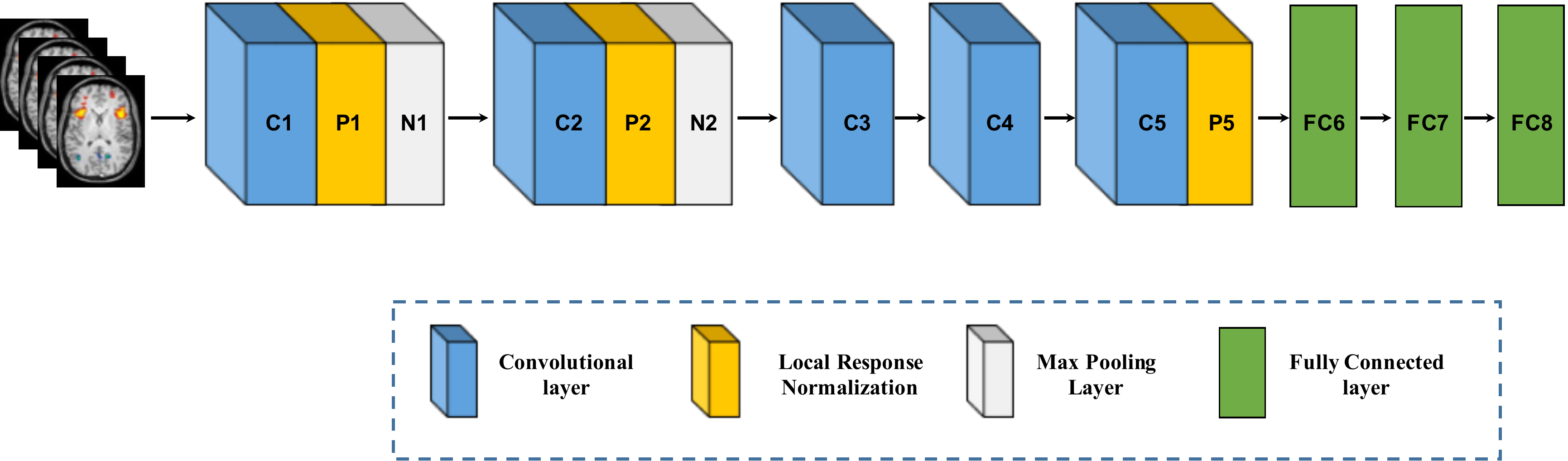}
	\caption{Architecture of convolutional neural networks}
	\label{Fig-CNN}
\end{figure*}

Compared to SAE, DBN and DBM,  utilizing the inputs in vector form which inevitably destroys the structural information in images, CNN is designed to better retain and utilize the structural information among neighboring pixels or voxels, and to required minimal preprocessing by directly taking two-dimensional(2D) or three-dimensional(3D) images as inputs~\cite{lecun1998gradient}.
Structurally, a CNN is a sequence of layers, and each layer of the CNN transforms one volume of activations to another through a differentiable function.
Figure~\ref{Fig-CNN} shows a typical CNN architecture for a computer vision task, which consists of three type neural layers: convolutional layers, pooling layers and fully connected layers. The convolutional layers interspersed with pooling layers, eventually leading to the fully connected layers.
The convolutional layer takes the pixels or voxels of a small patch of the input images, called the local receptive field, then utilizes various learnable kernels to convolve the receptive field to generate multiple feature maps.
A pooling layer performs the non-linear downsampling to reduce the spatial dimensions of the input volume for the next convolutional layer.
The fully connected layer input the 3D or 2D feature map to a 1D feature vector.

The major issue in training deep models is the over-fitting, which arises from the gap between the limited number of training samples and a large number of learnable parameters. Therefore, various techniques are designed to make the models train and generalize better, such as dropout and batch normalization to just name a few. A dropout layer randomly drops a fraction of the units or connections during each training iteration~\cite{srivastava2014dropout}. And it has been demonstrated that dropout is able to successfully avoid over-fitting. In addition, batch normalization is another useful regularization and performs normalization with the running average of the mean–variance statistics of each mini-batch.
It is shown that using batch normalization not only drastically speeds up the training time, but also improves the generalization performance~\cite{ioffe2015batch}.


\subsection{Graph Convolutional Networks}

\begin{figure*}[htbp]
	\centering
	\includegraphics[width=0.85\textwidth,angle=0]{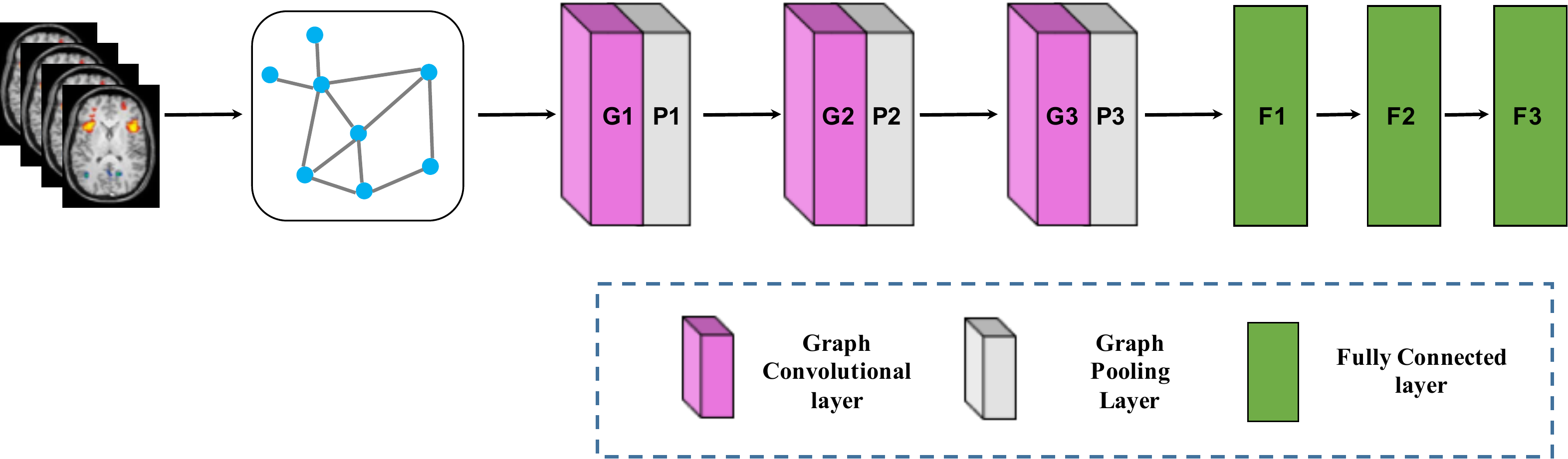}
	\caption{Architecture of graph convolutional  networks}
	\label{Fig-GCNN}
\end{figure*}

While CNN has achieved the huge success to extract latent representations from Euclidean data(\eg~images, text and video),
there are a rapidly increasing number of various applications where data is generated from the non-Euclidean domain and needs to be efficiently analyzed.
Researchers straightforwardly borrow ideas from CNN to design the architecture of graph convolutional networks (GCN) to handle complexity graph data~\cite{kipf2016semi}.
Figure~\ref{Fig-GCNN} shows the process of a simple GCN with graph pooling layers for a graph classification task.
The first step is to transform the traditional data to graph data, then the graph structure and node content information is regarded as input.
The graph convolutional layer plays a central role in extracting node hidden representations from aggregating the feature information from its neighbors.
The graph pooling layers can be interleaved with the GCN layers and coarsened graphs into sub-graphs in order to obtained higher graph-level representations for each node on coarsened sub-graphs.
After multiple fully connected layers, the softmax output layer is used to predict the class labels.

Depending on the types of graph convolutions, GCN can be categorized into spectral-based and spatial-based methods.
Spectral-based methods formulated graph convolution by introducing filters from the perspective of graph single processing.
Spatial-based methods defined graph convolution directly on the graph, which operates on spatial close neighbors to aggregate feature information.
Due to drawbacks to spectral-based methods from three aspects, efficiency, generality and flexibility, spatial-based methods have attracted more attention recently~\cite{wu2019comprehensive}.

\subsection{Recurrent Neural Networks}
\begin{figure*}[h]
	\centering
	\subfigure[Long Short-Term Memory ]{\includegraphics[width=0.42\textwidth,angle=0]{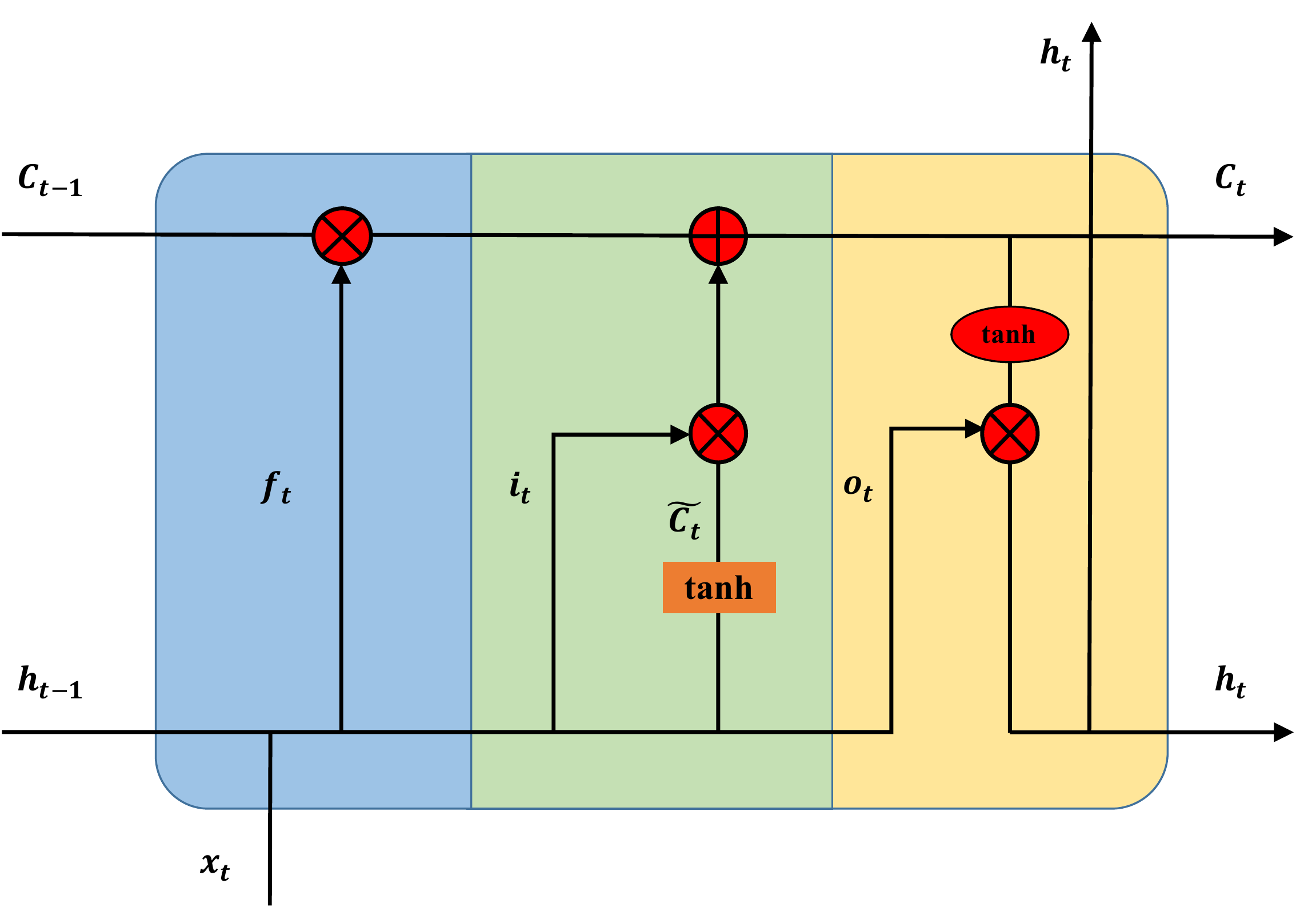}}
	\hfil
	\subfigure[Gated Recurrent Unit ]{\includegraphics[width=0.4\textwidth,angle=0]{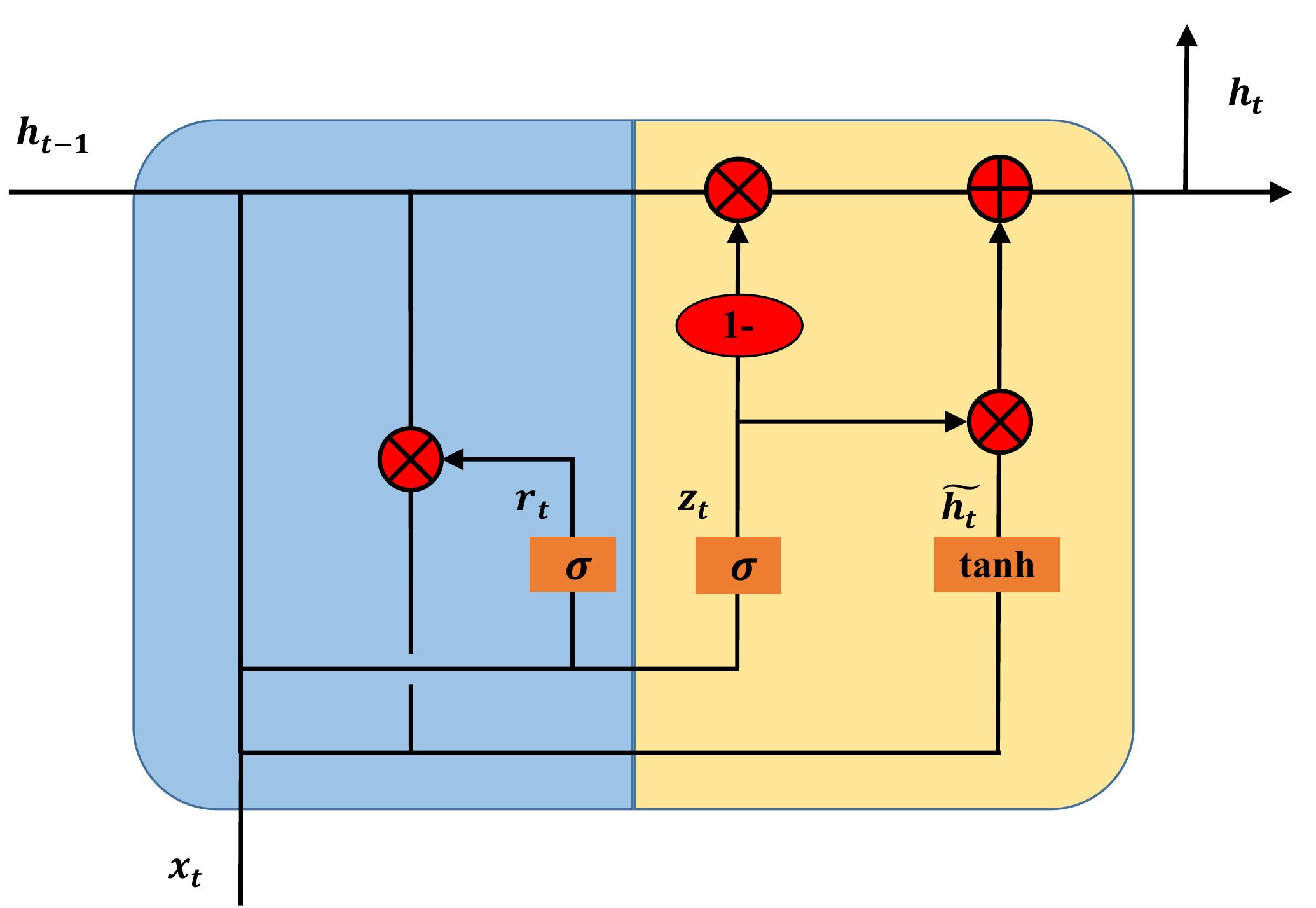}}
	\caption{Architectures of long short-term memory(a) and gated recurrent unit(b).
		In the subfigure (a), the blue, green and yellow represent the forget gate, input gate and output gate, respectively.
		In the subfigure (b), the blue and yellow represent the reset gate and update gate, respectively.
		 To keep the figure simple, biases are not shown.
	}
	\label{Fig-RNN}
\end{figure*}

A recurrent neural network (RNN) is an extension of an FFNN, which is able to learn features and long term dependencies from sequential and time-series data.
The most popular RNN architecture is the long-short term memory(LSTM)~\cite{hochreiter1997long},
which  is composed of a memory cell $C_{t}$,  a forget gate $f_t$, the input gate $i_t$ and the output gate $o_t$ (Figure~\ref{Fig-RNN}.a).
The memory cell transfers relevant information all the way to the sequence chain, and these gates control the activation singles from various sources to decide which information is added to and removed from the memory cell.

Unlike vanilla RNN, LSTM is able to decide whether to preserve the existing memory by the above-introduced gates. Theoretically, if LSTM learns an important feature from the input sequential data, it can keep this feature over a long time, thus captures potential long-time dependencies.
One popular LSTM variant is the Gated Recurrent Unit (GRU) (Figure~\ref{Fig-RNN}.b), which merges the forget and input gates into a single ``update gate'', and combines the memory cell state and hidden state into one state.  The update gate decides how much information to add and throw away, and the reset gate decides how much previous information to forget.
This makes the GRU is simpler than the standard LSTM~\cite{cho2014properties}.

\subsection{Open Source Deep Learning Library}
With the great successes of deep learning techniques in various applications, some famous research groups and companies have released their source codes and tools in deep learning. Due to these open source toolkits, people are able to easily build deep models for their applications even if they are not acquainted with deep learning technique.
Table~\ref{tab:software} lists the most popular toolkits for deep learning and shows their main features. 
All the software in the table can support for using GUP acceleration. 
For now, there are numerous deep learning toolkits available, but the problem it brings to people is how to select the most suitable toolkit for their applications.
Selecting the best toolkit depends on the goals of the projects, the characters of the available dataset, the skills and background of the researchers, the features of the available toolkits~\cite{erickson2017toolkits,zacharias2018survey}. Therefore, when a project starts, it is worth spending time to evaluate candidate toolkits to be sure that the best suitable toolkit is chosen for the corresponding application.

\begin{sidewaystable}
			\renewcommand\arraystretch{1.2}
	\caption{The popular open source toolkits for deep learning}
	\scriptsize
	\begin{tabular}{p{2.5cm}p{2.2cm}p{2cm}p{1.5cm}p{2.5cm}p{2cm}p{2.5cm}}	
		\hline
		\textbf{Name} & \textbf{Creator} & \textbf{GitHub} & \textbf{License} & \textbf{Platform} &\textbf{Language} & \textbf{Interface} \\ \hline 
		
		Caffe \cite{jia2014caffe} & Berkeley Center & BVLC/caffe & BSD & Linux, macOS, Windows & 	C++	 & Python, MATLAB, C++	 \\ \hline
		
		Deeplearning4j \cite{Deeplearning4j}& Skymind & deeplearning4j/ deeplearning4j &Apache 2.0 &Linux, macOS, Windows, Android & 	C++, Java	& Java, Scala, Clojure, Python,  Kotlin \\ \hline
		
		Keras \cite{Keras2015}& 	Franois Systems	& fchollet/keras   & MIT license	& Linux, macOS, Windows	& Python& 	Python, R
		\\ \hline
		
		MXNet \cite{chen2015mxnet}& 	Apache Software Foundation& apache/ incubator-mxnet	 & Apache 2.0	& Linux, macOS, Windows, AWS, Android, iOS, JavaScript&  C++	&C++, Python, Julia, Matlab, JavaScript, Go, R, Scala, Perl\\ \hline 
		
		TensorFlow \cite{Abadi2016TensorFlow}& 	Google & tensorflow/ tensorflow 	& Apache 2.0& 	 Linux, macOS, Windows, Android& 	C++, Python	& Python, C, C++, Java, Go, JavaScript, R, Julia, Swift
		\\ \hline 
		
		PyTorch \cite{paszke2017automatic}	& Adam Paszke et al& pytorch/ pytorch	 & BSD	&   Linux, macOS, Windows& 	Python, C 	& Python
		\\ \hline
		
		Theano \cite{AlRfou2016theano}& 	Université de Montréal	 & Theano/ Theano  & BSD& 	 Linux, macOS, Windows& 	Python& 	Python
		\\ \hline 
		
		Torch \cite{collobert2011torch}& 	Ronan Collobert et al & torch/ torch7	& BSD&  Linux, macOS, Windows, Android, iOS& 	C,  Lua, LuaJIT& C,Lua, LuaJIT \\ \hline 
		
		CNTK \cite{Seide2016cntk}& 	Microsoft	&  Microsoft/ CNTK &MIT license&  	 Linux, macOS, Windows & 	C++	&  Python, C++, C$\#$, Java\\ \hline 
		
		MATLAB & 	MathWorks & -		& Proprietary & Linux, macOS, Windows&C,C++, Java, MATLAB& 	MATLAB\\ \hline 
	\end{tabular}
	\label{tab:software}
\end{sidewaystable}

\section{Applications in Brain Disorder Analysis with Medical Images}
\label{S3}

The human brain is susceptible to many different disorders that strike at every stage of life.
Developmental disorders usually first appear in early childhood, such as autism spectrum disorder and dyslexia. Although psychiatric disorders are typically diagnosed in teens or early adulthood, their origins may exist much earlier in life, such as depression and schizophrenia. Then, as people age, people become increasingly susceptible to Alzheimer's disease, Parkinson's disease, and other dementia diseases.
In this section, we select four typical brain disorders, including Alzheimer's disease, Parkinson's disease, Autism spectrum disorder and Schizophrenia. Alzheimer's disease and Parkinson's disease are both neurodegenerative disorders. Autism spectrum disorder and Schizophrenia are neurodevelopmental and psychiatric disorders, respectively.

\subsection{Deep Learning for Alzheimer's Disease Analysis}

Alzheimer's disease (AD) is a neurological, irreversible, progressive brain disorder and is the most common cause of dementia.
Until now, the causes of AD are not yet fully understand, but accurate diagnosis of AD plays a significant role in patient care, especially at the early stage.
For the study of AD diagnosis, the best-known public neuroimaging dataset is from the Alzheimer's Disease Neuroimaging Initiative (ADNI), which is a multi-site study that aims to improve clinical trials for the prevention and treatment of AD. The ADNI study has been running since 2004 and is now in its third phase~\cite{mueller2005alzheimer}.
Researchers collect, validate and utilize data, including MRI and PET images, genetics, cognitive tests, cerebrospinal fluid (CSF) and blood biomarkers as predictors of the disease. Up to now, the ADNI dataset consists of ADNI-1, ADNI-GO, ADNI-2 and ADNI-3 and contains more than 1000 patients.
According to the Mini-Mental State Examination (MMSE) scores,  these patients were in three stages of disease: normal control (NC) , mild cognitive impairment(MCI) and AD.
The MCI subject can be divided into two subcategories: converted MCI (cMCI) and stable MCI (sMCI),  based on whether a subject converted to AD within a period of time (\eg~24 months).
The ADNI-GO and ADNI-2 provided two different MCI groups: early mild cognitive impairment (EMCI) and late mild cognitive impairment (LMCI), determined by a Wechsler Memory Scale (WMS) neuropsychological test.

Recently, plenty of papers have been published on the deep learning techniques for AD diagnosis.
According to different model architectures, these methods can be roughly divided into two subcategories: DGM-based and CNN-based methods. The DGM-based methods contained DBN, DNM, SAE and AE variants.
Li~\etal~stacked multiple RBM to construct a robust deep learning framework, which incorporated the stability selection and the multi-task learning strategy~\cite{li2015robust}.
Suk~\etal~proposed a series of methods based on deep learning models, such as DBM~\cite{suk2014hierarchical} and SAE~\cite{suk2015latent, suk2016deep}. For example, the literature~\cite{suk2015latent} applied SAE to learn the latent representations from sMRI, PET and CSF, respectively. Then, multi-kernel SVM classifier was used to fuse the selected multi-modal features.
Liu~\etal~also used SAE to extract features from multi-modal data, then a zero-masking strategy was applied to fuse these learned features~\cite{liu2015multimodal}.
Shi~\etal~adopted multi-modality stacked denoising sparse AE (SDAE) to fuse cross-sectional and longitudinal features estimated from MR brain images~\cite{shi2017nonlinear}.
Lu~\etal~developed a multiscale deep learning network, which took the multiscale patch-wise metabolism features as input~\cite{lu2018multiscale}.  And this study was perhaps the first study to utilize such a large number of FDG-PET images data.
Martinez-Murcia~\etal~used a deep convolution AE (DCAE)  architecture to extract features, which showed large correlations with clinical variables such as age, tau protein deposits, and especially neuropsychological examinations~\cite{martinez2019studying}.

CNN-based methods learned all levels of features from raw pixels and avoided the manual ROIs annotation procedure, and can be further subdivided into two subcategories: 2D-CNN and 3D-CNN.
Gupta~\etal~pre-trained a 2D-CNN based on sMRI data through a sparse AE on random patches of natural images~\cite{gupta2013natural}.
The key technique was the use of cross-domain features to present MRI data.
Liu~\etal~used a similar strategy and pre-trained a pre-trained deep CNN on ImageNet~\cite{liu2014learning}.
Sarraf~\etal~first used the fMRI data in deep learning applications~\cite{sarraf2016deepad}. The 4D rs-fMRI and 3D MRI data were decomposed into 2D  format images in the preprocessing step, and then the CNN-based architecture received these images in its input layer.
Billones~\etal~designed a DemNet model based on the 16-layers VGGNet. The DemNet only selected the coronal image slices with indices 111 to 130 in 2D format images under the assumption that these slices covered the areas, which had the important features for the classification task~\cite{billones2016demnet}.
Liu~\etal~proposed a novel classification framework that learned features from a sequence of 2D slices by decomposing 3D PET images~\cite{liu2018classification}. Then hierarchical 2D-CNN was built to capture the intra-slice features, while GRU was adopted to extract the inter-slice features.

\begin{table}[t]
		\renewcommand\arraystretch{1.3}
	\centering
	\scriptsize
	\caption{Overview of  papers using deep learning techniques for AD diagnosis.}
	
	\begin{tabular*}{1.2\textwidth}{@{\extracolsep{\fill}}lccccccccc}
		\hline
		\multirow{2}{*}{Reference}&  \multirow{2}{*}{Year} & \multirow{2}{*}{Database} &
		\multicolumn{4}{c}{Subjects}& \multirow{2}{*}{Modality} &\multirow{2}{*}{Model}  \\\cline{4-7}
		
		& & &AD&cMCI&sMCI&NC&& \\
		\hline
				Li~\etal~\cite{li2015robust} & 2015 & ADNI & 51 & 43 &56 & 52 & sMRI+PET+CSF & DBN \\\hline
		Liu~\etal~\cite{liu2015multimodal} & 2015 & ADNI &  85  & 67 & 102 & 77  & sMRI+PET & SAE \\\hline

		Suk~\etal~\cite{suk2014hierarchical}.  & 2014 &ADNI &   93 & 76 & 128 & 101 & sMRI+PET & DBM \\\hline
		Suk~\etal~\cite{suk2015latent}  & 2015& ADNI   &     51 & 43 & 56 & 52 & sMRI+PET+CSF & SAE   \\  \hline
		\multirow{2}{*}{Suk~\etal~\cite{suk2016deep} }  &   \multirow{2}{*}{2016}  & ADNI &51 & 43 & 56 & 52&  \multirow{2}{*}{sMRI+PET+CSF} &  \multirow{2}{*}{SAE}   \\\cline{4-7}
		& & - &198 & 167& 236 &229 &     \\\hline		
		Shi~\etal~\cite{shi2017nonlinear}  & 2016 &ADNI & 95  & \multicolumn{2}{c}{121} &123 &  sMRI + Age &   SDAE     \\		\hline		
		Lu~\etal~\cite{lu2018multiscale} & 2018 &   ADNI&   226& 112& 409 & 304 &PET & SAE  \\  \hline
		Martinez-Murcia~\etal~\cite{martinez2019studying}& 2019 &ADNI & 99&  \multicolumn{2}{c}{212} &168&rs-fMRI&DCAE \\\hline

		Gupta~\etal~\cite{gupta2013natural} & 2013& ADNI & 200 & \multicolumn{2}{c}{411} & 232 &sMRI & 2D-CNN  \\	\hline	
		Liu~\etal~\cite{liu2014learning} & 2014& ADNI & 200 & \multicolumn{2}{c}{411} & 232  & sMRI & 2D-CNN  \\\hline
		Billones~\etal~\cite{billones2016demnet} &2016& ADNI & 300 & \multicolumn{2}{c}{300} & 300  &rs-fMRI &2D-CNN    \\		\hline
		\multirow{2}{*}{Sarraf~\etal~\cite{sarraf2016deepad} }&\multirow{2}{*}{2016} & \multirow{2}{*}{ADNI} & 211 & - &- & 91&sMRI & \multirow{2}{*}{2D-CNN} &\\\cline{4-7}
		& & & 52 &-&-& 92 & rs-fMRI &	\\ \hline
		Liu~\etal~\cite{liu2018classification} & 2017 & ADNI&  93& \multicolumn{2}{c}{146} & 100 &  PET & 2D-CNN+RNN  \\			\hline

		Payan~\etal~\cite{payan2015predicting}&2015& ADNI & 755 & \multicolumn{2}{c}{755} & 755  & sMRI & 3D-CNN  \\	\hline
		Hosseini-Asl~\etal~\cite{hosseini2016alzheimer}&2016& ADNI & 70 & \multicolumn{2}{c}{70} & 70  & sMRI& 3D-CNN \\	\hline		
		Karasawa~\etal~\cite{karasawa2018deep} &2017& ADNI & 348 & 450& 358  & 574  & sMRI &3D-CNN\\	\hline
		Liu~\etal~\cite{liu2018multi} & 2018 & ADNI & 93 & 76 &128  & 100  & sMRI+PET &  	3D-CNN \\\hline		
		Li~\etal~\cite{li2014deep} &2014& ADNI & 193 & 167 & 236 & 229  & sMRI+ PET &3D-CNN \\\hline
		Liu~\etal~\cite{liu2018landmark}&2018& ADNI & 358 &205 & 465 & 429  & sMRI &3D-CNN \\\hline
		Liu~\etal~\cite{liu2018anatomical}&2018& ADNI & 358 &- & - & 429    & sMRI &3D-CNN \\\hline
		Pan~\etal~\cite{Pan2018Synthesizing} &2018& ADNI & 358 &205 & 465 & 429   & sMRI+ PET &3D-CNN+GAN \\
		\hline	
		
	\end{tabular*}
	\label{tab:AD01}
\end{table}

Because the 3D brain images need to be decomposed into 2D slices in the preprocessing step,  it resulted in 2D-CNN methods discarded the spatial information. Therefore, many 3D-CNN methods were proposed, which can directly input 3D brain images.
Payan~\etal~was pre-trained a 3D-CNN through a sparse AE on small 3D patches from sMRI scans~\cite{payan2015predicting}.
Hosseini-Asl~\etal~ proposed a deep 3D-CNN,  which was built upon a 3D CAE to capture anatomical shape variations in sMRI scans~\cite{hosseini2016alzheimer}. Liu~\etal~used multiple deep 3D-CNN on different local image patches to learn the discriminative features of MRI and PET images.
Then a set of upper high-level CNN was cascaded to ensemble the learned local features and discovered the latent multi-modal features for AD classification~\cite{liu2018multi}.
Karasawa~\etal~proposed deeper 3D-CNN architecture with 39 layers based on residual learning framework(ResNet) to improve performance~\cite{karasawa2018deep}.
Liu~\etal~designed a  landmark-based deep feature learning framework to learn the patch-level features, which were an intermediate scale between voxel-level and ROI-level~\cite{liu2018anatomical}. The authors firstly used a data-driven manner to identify discriminative anatomical landmarks from MR images, and then proposed a 3D-CNN to learn patch-based features. This strategy can avoid the high-dimensional problem of voxel-level and manual definition of ROI-level.
Subsequently, Liu~\etal~developed a deep multi-instance CNN framework~\cite{liu2018landmark}, where multiple image patches were used as a bag of instances to represent each specific subject, and then the label of each bag was given by the whole-image-level class label.
To overcome the missing modality in multi-modal image data, Li~\etal~\cite{li2014deep} proposed a simple 3D-CNN to predict the missing PET images from the sMRI data. Results showed that the predicted PET data achieved similar classification accuracy as the true PET data. Additionally,  the synthetic PET data and the real sMRI data obviously outperformed the single sMRI data.
Pan~\etal~used Cycle-GAN to learn bi-directional mapping sMRI and PET, to synthesize missing PET scans based on its corresponding sMRI scans. Then, landmark-based 3D-CNN was adapted for AD classification on the mixed image data~\cite{Pan2018Synthesizing}.
Table \ref{tab:AD01} and Table \ref{tab:AD02}  summarized the statistic information of each paper reviewed above for AD diagnosis.

\begin{table}[t]
	\begin{threeparttable}
				\renewcommand\arraystretch{1.3}
		\centering
		\scriptsize
		\caption{The classification performance of papers  for AD diagnosis.}
		
		\begin{tabular*}{1.2\textwidth}{@{\extracolsep{\fill}}lcccccc}
			\hline
			\multirow{2}{*}{Reference}& \multicolumn{6}{c}{ Accuracy$(\%)$}\cr \cline{2-7}
			
			& AD/NC&AD/MCI& MCI/NC  & cMCI/sMCI & 3-ways\tnote{1} & 4-ways\tnote{2}\\
			\hline
				Li~\etal~\cite{li2015robust}  & 91.4$\pm$ 1.8 &  70.1$\pm$ 2.3&77.4$\pm$ 1.7& 57.4$\pm$ 3.6\\\hline		
			Liu~\etal~\cite{liu2015multimodal} &  91.4$\pm$ 5.56 &-&  82.10$\pm$ 4.91 & -  & -  & 53.79\\\hline

			Suk~\etal~\cite{suk2014hierarchical}. & 95.35 $\pm$ 5.23 &- & 85.67$\pm$ 5.22 &  75.92$\pm$ 15.37 & -  & -  \\\hline		

			Suk~\etal~\cite{suk2015latent}  &98.8 $\pm$0.9 & 83.7$\pm$1.5 & 90.7$\pm$1.2 & 83.3 $\pm$2.1 &-&- \\  \hline
			\multirow{2}{*}{Suk~\etal~\cite{suk2016deep} } & 95.09$\pm$ 2.28& - &80.11 $\pm$ 2.64  &74.15 $\pm$ 3.35&62.93&53.72 \\\cline{2-7}
			&  90.27   & - & 70.86 & 73.93 & 57.74 & 47.83   \\\hline

			Shi~\etal~\cite{shi2017nonlinear}  & 91.95 $\pm$1.00  & - & 83.72$\pm$ 1.16 & - & - & -  \\		\hline
			Lu~\etal~\cite{lu2018multiscale} &   93.58  $\pm$ 5.2 &- &-& 81.55 $\pm$7.42 & - &-\\  \hline
			Martinez-Murcia~\etal~\cite{martinez2019studying}& 84.3$\pm$6 & -& - & 71.5$\pm$9 &- &-   \\\hline

			Gupta~\etal~\cite{gupta2013natural} &94.74 & 88.10 & 86.35 & - & 85.0 & -  \\		\hline
			Liu~\etal~\cite{liu2014learning} & 97.18$\pm$1.5 & 94.51$\pm$1.43 & 93.21$\pm$1.02 &- & 91.72$\pm$1.8&-\\\hline
			Billones ~\etal~\cite{billones2016demnet} &98.33 & 93.89 & 91.67& -& 91.85&-  \\		\hline

			Sarraf~\etal~\cite{sarraf2016deepad} & 98.84 /99.90 & - &- & - &-&-	\\ \hline
			Liu ~\etal~\cite{liu2018classification} &91.92   &  -&  78.9  &-&-&- \\\hline
			
			Payan~\etal~\cite{payan2015predicting}& 95.39& 86.84&92.11&- & 89.47 &-\\	\hline
			Hosseini-Asl~\etal~\cite{hosseini2016alzheimer}& 99.3$\pm$1.6&100& 94.2$\pm$2.0&-& 94.8$\pm$2.6&- \\	\hline
			Karasawa~\etal~\cite{karasawa2018deep} &94.0 & -& 90.0 &-& 87.0&-\\	\hline

			Liu~\etal~\cite{liu2018multi} & 93.26& - &73.34 &-&- &-\\\hline

			Li~\etal~\cite{li2014deep} &92.87$\pm$2.07 & -& 76.21$\pm$2.05 &72.44$\pm$2.41&-&-\\\hline
			Liu~\etal~\cite{liu2018landmark}&91.09 &-&- &76.90 &-&-\\\hline
			Liu~\etal~\cite{liu2018anatomical}&90.56 &-&- &- &-&-\\\hline
			Pan~\etal~\cite{Pan2018Synthesizing} &92.50 &-&- &79.06 &-&-\\
			\hline	
			
		\end{tabular*}
		\begin{tablenotes}
			\item[1] 3-ways  represents the comparison: AD vs. NC vs. MCI.
			\item[2] 4-ways represents the comparison: AD vs. NC vs. cMCI vs. sMCI.

		\end{tablenotes}
		
		\label{tab:AD02}
	\end{threeparttable}
\end{table}

As an early stage of AD, MCI had a conversion rate as high as 10\%-15\% per year in 5 years, but MCI was also the best time for treatment. Therefore, an effective predictive model construction for the early diagnosis of MCI had become a hot topic. Recently, some research based on GCN has been done for MCI prediction.
Zhao~\etal~and Yu~\etal~both used GCN which combines neuroimaging information and demographic relationship for MCI prediction~\cite{zhao2019graph,yu2019multi}.
Song~\etal~ implemented a multi-class GCN classifier for classification of subjects on the AD spectrum into four classes~\cite{song2019graph}.
Guo~\etal~proposed PETNET model based on GCN to analyzes PET signals defined on a group-wise inferred graph structure~\cite{guo2019predicting}. Table \ref{tab:AD03} and Table \ref{tab:AD04} summarized the four papers for MCI prediction.

\begin{table}[!htbp]
		\renewcommand\arraystretch{1.3}
	\centering
	\scriptsize
	\caption{Overview of  papers using deep learning techniques for MCI prediction.}
	
	\begin{tabular*}{1.2\textwidth}{@{\extracolsep{\fill}}lccccccccc}
		\hline
		\multirow{2}{*}{Reference}&  \multirow{2}{*}{Year} & \multirow{2}{*}{Database} &
		\multicolumn{4}{c}{Subjects}& \multirow{2}{*}{Modality} &\multirow{2}{*}{Model}  \\\cline{4-7}
		
				& & &NC&EMCI&LMCI&AD&& \\ \hline

	Zhao~\etal~\cite{zhao2019graph} & 2019 & ADNI & 67 & 77 &40 &- & rs-fMRI& GCN \\\hline
Yu~\etal~\cite{yu2019multi}    & 2019 & ADNI & 44 & 44 &38 &- & rs-fMRI& GCN \\\hline
		Song~\etal~\cite{song2019graph} & 2019 & ADNI & 12 & 12 &12 &12 & DTI& GCN \\\hline
		Guo~\etal~\cite{guo2019predicting} & 2019 & ADNI & 100 & 96 &137 &- & PET& GCN \\\hline

	\end{tabular*}
\label{tab:AD03}
\end{table}

\begin{table}[!htbp]
\begin{threeparttable}
			\renewcommand\arraystretch{1.3}
	\centering
	\scriptsize
	\caption{The classification performance of papers for MCI prediction.}  	
	\begin{tabular*}{1.2\textwidth}{@{\extracolsep{\fill}}lcccccc}
	\hline
	\multirow{2}{*}{Reference}& \multicolumn{6}{c}{ Accuracy$(\%)$}\cr \cline{2-7}

	&EMCI/NC&LMCI/NC& EMCI/LMIC & MCI/NC&3-ways\tnote{1} & 4-ways\tnote{2}\\\hline
		
			Zhao~\etal~\cite{zhao2019graph} & 78.4 & 84.3 & 85.6&-&-&-  \\\hline
			Yu~\etal~\cite{yu2019multi} &87.5& 89.02 & 79.27&-&-&-   \\\hline
		Song~\etal~\cite{song2019graph} &-& - & -&-&-&89.0$\pm$6\\\hline

		Guo~\etal~\cite{guo2019predicting} & - & -& -&93.0\tnote{3}&77.0&-  \\\hline

	\end{tabular*}

		\begin{tablenotes}
	\item[1] 3-ways  represents the comparison: NC vs. EMCI vs. LMCI.
	\item[2] 4-ways represents the comparison: NC vs. EMCI vs. LMCI vs. AD.
	\item[3] MCI = ECMI+LMCI
\end{tablenotes}

	\label{tab:AD04}
\end{threeparttable}

\end{table}

\subsection{Deep Learning for Parkinson's Disease Analysis}

Parkinson's disease (PD) is the most common neurodegenerative disorder, after Alzheimer's disease, and is provoked by progressive impairment and deterioration of neurons, caused by a gradually halt in the production of a chemical messenger in the brain.
Parkinson's Progression Markers Initiative (PPMI) is an observational clinical study to verify progression markers in Parkinson's disease.
The PPMI cohort comprises 400 newly diagnosed PD cases, 200 healthy, and 70 individuals, while clinically diagnosed as PD cases, fail to show evidence of dopaminergic deficit. This latter group of patients is referred to as SWEDDs (Scans without Evidence of Dopamine Deficit)~\cite{marek2011parkinson}.

Some efforts based on deep learning have been done to designed algorithms to help PD diagnosis.
Martinez-Murci team has continuously published a series of papers using deep learning techniques for PD diagnose in SPECT image dataset.
Ortiz~\etal~designed a framework to automatically diagnose PD using deep sparse filtering-based features~\cite{ortiz2016automated}. Sparse filtering, based on $\ell_2$-norm regularization, extracted the suitable features,  which can be used as the weight of hidden layers in a three-layer DNN.
Subsequently, this team firstly applied 3D-CNN in PD diagnosis. These methods achieved up to a 95.5$\%$ accuracy and 96.2$\%$ sensitively~\cite{martinez20173d}. However, this 3D-CNN architecture with only two convolutional layers was too shallow and limited the capability to extract more discriminative features.
Therefore, Martinez-Murcia~\etal~proposed a deep convolutional AE (DCAE) architecture for feature extraction~\cite{martinez2018deep}. The DCAE overcome two common problems: the need for spatial normalization and the effect of imbalanced datasets. For a strongly imbalanced (5.69/1) PD dataset, DCAE achieved more than 93$\%$ accuracy.
Choi~\etal~developed a deep CNN model (PDNet) consisted of four 3D convolutional layers~\cite{choi2017refining}.
PDNet obtained high classification accuracy compared to the quantitative results of expert assessment, and can further classify the SWEDD and NC subjects.
Esmaeilzadeh~\etal~both utilized the sMRI scans and demographic information (\ie, age and gender) of patients to train a 3D-CNN model~\cite{esmaeilzadeh2018end}.
The proposed method firstly found that the $Superior$ $Parietal$ part on the right hemisphere of the brain was critical in PD diagnosis.
Sivaranjini~\etal~directly introduced the AlexNet model, which was trained by the  transfer learned network~\cite{sivaranjini2019deep}.
Shen~\etal~proposed an improved DBN model with an overlapping group lasso sparse penalty to learn useful low-level feature representations~\cite{shen2019use}.
To incorporate multiple brain neuroimaging modalities, McDaniel~\etal~and Zheng~\etal~both used a GCN model and presented an end-to-end pipeline  without extra parameters involved for view pooling and pairwise matching~\cite{mcdaniel2019developing,zhang2018multi}.
Table \ref{tab:PD01} summarized each paper above reviewed  for PD diagnosis.

\begin{table}[!htbp]
	\begin{threeparttable}
		\renewcommand\arraystretch{1.3}
		\centering
		\scriptsize
		\caption{Overview of papers using deep learning techniques for PD diagnose.}
		\begin{tabular*}{1.35\textwidth}{@{\extracolsep{\fill}}lccccccccc}
			\hline
			\multirow{2}{*}{Reference} & Year &   Database &Modality  & Method & \multicolumn{3}{c}{Modality}  &  \multicolumn{2}{c}{Accuracy$(\%)$}  \\\cline{6-8}\cline{9-10}
			&&&&	& PD & NC  &SWEED  & PD/NC   &SWEED/NC \\\hline
			
			Ortiz~\etal~\cite{ortiz2016automated}  & 2016 &    PPMI&  SPECT   &DNN  & -&-&-& 95.0 &-  \\		\hline
			Martinez-Murcia~\etal~\cite{martinez20173d} & 2017 &   PPMI&  SPECT    &3D-CNN& 158&111&32& 95.5$\pm$4.4&82.0$\pm$6.8   \\		\hline
			\multirow{2}{*}{Choi~\etal~\cite{choi2017refining}}& \multirow{2}{*}{2017} & PPMI&SPECT & 	\multirow{2}{*}{3D-CNN}  &431&193&77& 96.0 &76.5\\\cline{6-10}\cline{3-4}
			&     & SNUH\tnote{1} &SPECT &	&  72&10& - &  98.8& -  \\		\hline
			Esmaeilzadeh~\etal~\cite{esmaeilzadeh2018end}& 2018 &  PPMI&  sMRI+DI\tnote{4} &3D-CNN&452 & 204&- &1.0 &- \\		\hline
			Martinez-Murcia~\etal~\cite{martinez2018deep} & 2018 & PPMI&  SPECT    &DCAE    & 1110 & 195 & - & 93.3$\pm$1.6 &-    \\			
			\hline
			Sivaranjini~\etal~\cite{sivaranjini2019deep} & 2019& PPMI & SPECT & 2D-CNN& 100 & 82 & - & 88.9& - \\\hline
			McDaniel~\etal~\cite{mcdaniel2019developing} & 2019 & PPMI & sMRI+DTI& GCNN & 117&30 &- & 92.14& - \\\hline
			Zhang~\etal~\cite{zhang2018multi}& 2018 & PPMI & sMRI+DTI & GCNN & 596 & 158 & - & 95.37(AUC)&-\\\hline
			\multirow{2}{*}{Shen~\etal~\cite{shen2019use}}& \multirow{2}{*}{2019} & HSHU\tnote{2}&PET & 	\multirow{2}{*}{DBN}  &100&200&-& 90.0 &-\\\cline{6-10}\cline{3-4}
                              &     & WXH\tnote{3} &PET &	&  25&25& - & 86.0& -  \\		\hline
	 			\end{tabular*}
		\begin{tablenotes}
			\item[1] SNUH: Seoul National University Hospital cohort.
			\item[2] HSH: HuaShan Hospital cohort.
			\item[3] WXH: WuXi 904 Hospital cohort.
			\item[4] DI: Demographic Information.
		\end{tablenotes}
		\label{tab:PD01}
	\end{threeparttable}
\end{table}

\subsection{Deep Learning for Austism Spectrum Disorder Analysis}

Autism spectrum disorder (ASD) is a common neurodevelopmental disorder, which has affected 62.2 million ASD cases in the world in 2015.
The Autism Imaging Data Exchange (ABIDE) initiative had aggregated rs-fMRI brain scans, anatomical and phenotypic datasets, collected from laboratories around the world.
The ABIDE initiative included two large scale collections: ABIDE I and ABIDE II, which were released in 2012 and 2016, respectively. The ABIDE I collection involved 17 international sites, and consisted of 1,112 subjects comprised of 539 from autism patients and 573 from NC. To further enlarge the number of samples with better-characterized, the ABIDE II collection involved 19 international sites, and aggregated 1114 subjects from 521 individuals with ASD and 593 NC subjects~\cite{Di2014The}.

Many methods have been proposed on the application of deep learning for ASD diagnosis.
These methods can be divided into three categories: AE-based methods, convolutional-based methods, and RNN-based methods.
AE-based methods used various AE variations or stacked multiple AE to reduce data dimension and discovery highly discriminative representations.
Hazlett~\etal~implemented the basic SAE which primarily used surface area information from brain MRI at 6 and 12 month old infants to predict the 24 month diagnosis of autism in children at high familial risk for autism. The SAE contained 3 hidden layers to reduce 315 dimension measurements to only 2 features~\cite{hazlett2017early}.
Two papers both used a stacked multiple sparse AE (SSAE) to learn low dimensional high-quality representations of functional connectivity patterns~\cite{Guo2017Diagnosing, kong2019classification}. But the difference was that Guo~\etal~input the whole-brain functional connectivity patterns and Kong~\etal~only selected top 3,000 ranked connectivity features by F-score in descending order.
Dekhil~\etal~built an automated autism diagnosis system, which used 34 sparse AE for 34 spatial activation areas respectively~\cite{dekhil2018using}. Each sparse AE extracted the power spectral densities (PSDs) of time courses in a higher level representation and simultaneously reduced the feature vectors dimensionality.
Choi~\etal~used VAE to summarize the functional connectivity networks into two-dimensional features~\cite{choi2017functional}. One feature was identified  with a highly discrimination between ASD and NC, and  was closely associated with  ASD-related brain regions.
Heinsfeld~\etal~used DAE to reduce the effect of multi-site heterogeneous data and improve the generalization~\cite{Heinsfeld2018Identification}.
Due to insufficient training samples, Li~\etal~developed a novel deep neural network framework with the transfer learning technique for enhancing ASD classification~\cite{li2018novel}. This framework firstly trained an SSAE to learn functional connectivity patterns from healthy subjects in the existing databases. Then the trained SSAE was transferred to a new classification with limited target subjects.
Saeed~\etal~designed a data augmentation strategy to produce synthetic datasets needed for training the ASD-DiagNet model. This model was composed of an AE and a single-layer perceptron to improve the quality of extracted features~\cite{saeed2019asd}.

Due to collapsed the rs-fMRI scans into a  feature vector, above methods discarded the spatial structure of the brain networks.
To fully utilize the whole brain spatial fMRI information, Li~\etal~implemented 3D-CNN to capture spatial structure information and used sliding windows over time to measure temporal statistics~\cite{li20182}. This model was able to learn ASD related biological markers from the output of the middle convolution layer.
Khosla~\etal~proposed a 3D-CNN framework for connectome-based classification. The functional connectivity of each voxel to various target ROIs was used as input features, which reserved the spatial relationship between voxels. Then the ensemble learning strategy was employed to average the different ROI definitions to reduce the effect of empirical selections, and obtained more robust and accurate results~\cite{khosla20183d}.
Ktena~\etal~implemented a siamese GCN to learn a graph similarity metric, which took the graph structure into consideration for the similarity between a pair of graphs~\cite{ktena2018metric}. This was the first application of metric learning with graph convolutions on brain connectivity networks.
Parisot~\etal~introduced a spectral GCN for brain analysis in populations combining imaging and non-imaging information~\cite{parisot2017spectral}. The populations were represented as a sparse graph where each vertex corresponded to an imaging feature vector of a subject and the edge weights were associated with phenotypic data, such as age, gender, acquisition sites. Like the graph-based label propagation, a GCN model was used to infer the classes of unlabelled nodes on the partially labeled graphs.
There existed no definitive method to construct reliable graphs in practice. Thus, Anirudh \etal~proposed a bootstrapped version of GCN to reduce the sensitivity of models on the initial graph construction step~\cite{anirudh2017bootstrapping}. The bootstrapped GCN used an ensemble of weakly GCN, each of which was trained by a random graph.
In addition, Yao~\etal~proposed a multi-scale triplet GCN to avoid the spatial limitation of a single template~\cite{yao2019triplet}. A multi-scale templates for coarse-to-fine ROI parcellation were applied to construct multi-scale functional connectivity patterns for each subject. Then a triple GCN model was developed to learn multi-scale graph features of brain networks.

\begin{table}[t]
		\renewcommand\arraystretch{1.3}
	\centering
	\scriptsize
	\caption{Overview of papers using deep learning techniques for ASD diagnosis.}
	\begin{threeparttable}
		\begin{tabular*}{1.2\textwidth}{@{\extracolsep{\fill}}lcccccccc}
			\hline
			\multirow{2}{*}{Reference} & \multirow{2}{*}{Year} & \multirow{2}{*}{Database} & \multicolumn{2}{c}{Subject}   & \multirow{2}{*}{Modality} & \multirow{2}{*}{Model}  & \multirow{2}{*}{Accuracy$(\%)$} \\\cline{4-5}
			&&&ASD&NC&  \\	\hline
			Guo~\etal~\cite{Guo2017Diagnosing}& 2017 &ABIDE I & 55 &55 &rs-fMRI & SSAE  &86.36 \\	\hline
			Kong~\etal~\cite{kong2019classification}& 2019&ABIDE I & 78 & 104 &rs-fMRI & SSAE & 90.39\\	\hline
			\multirow{4}{*}{Li~\etal~\cite{li2018novel} }&\multirow{4}{*}{2018 }  & ABIDE: UM\tnote{1} &48&65 & \multirow{4}{*}{ rs-fMRI}& \multirow{4}{*}{SSAE}&67.2\\\cline{3-5}
			&  &   ABIDE:UCLA\tnote{2}  & 36&39 & && 62.3 \\\cline{3-5}
			&  &  ABIDE: USM\tnote{3} &38&23&&& 70.4\\\cline{3-5}
			&  &  ABIDE: LEUVEN\tnote{4}&27&34&&&68.3\\\hline
			
			Choi~\etal~\cite{choi2017functional} &2017 & ABIDE  & 465 & 507 &rs-fMRI& VAE& 0.60(AUC)\\\hline
			Heinsfeld~\etal~\cite{Heinsfeld2018Identification}& 2018 & ABIDE &505 &530&rs-fMRI &DAE & 70.0 \\\hline
			Hazlett~\etal~\cite{hazlett2017early}&2017&NDAR\tnote{5}&106&42&rs-fMRI & SAE & 88.0\\\hline
			Dekhil~\etal~\cite{dekhil2018using}&2018&NDAR&123 & 160& rs-fMRI&SSAE & 91.0$\pm$3.2\\\hline
			Saeed~\etal~\cite{saeed2019asd} &2019& ABIDE& 505 & 530 & rs-fMRI & AE & 70.1$\pm$3.2 \\\hline
			Li~\etal~\cite{li20182}& 2018& -& 82 & 48&rs-fMRI & 3D-CNN &89.0$\pm$5.0(F-score)\\\hline

			Khosla~\etal~\cite{khosla20183d} &2018& ABIDE&542 & 625 & rs-fMRI & 3D-CNN& 73.3\\\hline
			
			Ktena~\etal~\cite{ktena2018metric}& 2017 &ABIDE&403&468&rs-fMRI &GCN &62.9\\\hline
			Parisot~\etal~\cite{parisot2017spectral}& 2017 &ABIDE&403&468&rs-fMRI &GCN & 69.5 \\\hline
			Anirudh~\etal~\cite{anirudh2017bootstrapping}& 2017&ABIDE&404&468 &rs-fMRI &GCN &70.8\\\hline
			Yao~\etal~\cite{yao2019triplet} &2019& ABIDE& 438&544 & rs-fMRI& GCN& 67.3 \\\hline
			Bi~\etal~\cite{bi2018diagnosis} &2018&ABIDE&50&42&rs-fMRI& RNN  &84.7$\pm$3.2\\\hline
			Dvornek~\etal~\cite{dvornek2017identifying}&2017&ABIDE& 1100 & -&rs-fMRI& LSTM&68.5$\pm$5.5\\\hline

			\hline
		\end{tabular*}
		\begin{tablenotes}
			\item[1] University of Michigan.
			\item[2] University of California, Los Angeles.
			\item[3] University of Utah School of Medicine.
			\item[4] Katholieke Universiteit Leuven.
			\item[5] National Database of Autism Research.
		\end{tablenotes}
		
		\label{tab:ASD01}
	\end{threeparttable}
\end{table}

Several RNN-based methods were proposed to fully utilize the temporal information in the rs-fMRI time-series data.
Bi~\etal~designed a random NN cluster, which combined multiple NNs into a model, to improve the classification performance in the diagnosis of ASD~\cite{bi2018diagnosis}. Compared to five different NNs, the random Elman cluster obtained the highest accuracy. It is because that the Elman NN fit handling the dynamic data.
Dvornek~\etal~first applied LSTM to ASD classification, which directly used the rs-fMRI time-series data, rather than the pre-calculated measures of brain functional connectively~\cite{dvornek2017identifying}. The authors thought that the rs-fMRI time-series data contained more useful information of dynamic brain activity than single and static functional connectivity measures.
For clarity, the important information of the above-mentioned papers was summarized in Table~\ref{tab:ASD01}.

\subsection{Deep Learning for Schizophrenia Analysis}

Schizophrenia (SZ) is a prevalent psychiatric disorder and affects 1\% of the population worldwide. Due to the complex clinical symptoms, the pathological mechanism of schizophrenia remains unclear and there is no definitive standard in the diagnosis of SZ.
Different from the ADNI for AD diagnosis, the PPMI for PD diagnosis and the ABIDE  for ASD diagnosis,
there was not a widely used neuroimaging dataset for the SZ diagnosis. Therefore, many studies utilized various source datasets that were available from the medical research centers, universities and hospitals.

Recently, some studies have been successfully applied deep learning algorithms to SZ diagnosis and received significant improvement.
These methods were divided into two categories: unimodality and multi-modality, according to the types of input data, rather than according to deep learning architectures like AD or ASD diagnosis.

The unimodality category only used a single type of MRI, and can further be classified into subclasses: sMRI-methods and fMRI-methods. sMRI-methods discovery latent features from sMRI dataset, which can provide information on the tissue structure of the brain, such as gray matter, white matter and cerebrospinal fluid.
Plis~\etal~and Pinaya~\etal~used the DBN model, which only contained three hidden layers, to automatically extract feature for SZ identification. The results achieved modestly higher predictive performance than the shallow-architecture SVM approach~\cite{Plis2014Deep,Pinaya2016Using}.
Different from DBN model in~\cite{Pinaya2016Using}, Pinaya~\etal~trained an SAE to create a normative model from 1113 NC subjects, then used this model to estimate total and regional neuroanatomical deviation in individual patients with SZ~\cite{pinaya2019using}.
Ulloa~\etal~proposed a novel classification architecture that used synthetic sMRI scans to mitigate the effects of a limited sample size. To generate synthetic samples, a data-driven simulator was designed that can capture statistical properties from observed data using independent component analysis (ICA) and a random variable sampling method. Then a 10-layer DNN was trained exclusively on continuously generated synthetic data, and greatly improves generalization in the classification of SZ patients and NC~\cite{Ulloa2015Synthetic}.

\begin{table}[t]
		\renewcommand\arraystretch{1.3}
	\centering
	\scriptsize
	\caption{Overview of papers using deep learning techniques for SZ diagnosis.}
	\begin{threeparttable}
		\begin{tabular*}{1.3\textwidth}{@{\extracolsep{\fill}}lcccccccc}
			\hline
			\multirow{2}{*}{Reference} & \multirow{2}{*}{Year} & \multirow{2}{*}{Database} & \multicolumn{2}{c}{Subject}   & \multirow{2}{*}{Modality} & \multirow{2}{*}{Model}  & \multirow{2}{*}{Accuracy$(\%)$} \\\cline{4-5}
			&&&SZ&NC&  \\	\hline
			Plis~\etal~\cite{Plis2014Deep}& 2014 &Multi-site1\tnote{1} &198&191 &sMRI&DBN& 91.0+14(F-score) \\\hline
			Ulloa~\etal~\cite{Ulloa2015Synthetic}& 2015 &Multi-site1&198&191&sMRI &DNN & 75.0$\pm$4(AUC)\\\hline
			Pinaya~\etal~\cite{Pinaya2016Using}& 2016 &UNIFESP\tnote{2}&143&83&sMRI &DBN&73.55$\pm$6.84 \\\hline
			Pinaya~\etal~\cite{pinaya2019using}&2019 &NUSDAST\tnote{3} & 30&40& sMRI &SAE&70.7\\\hline

			Kim~\etal~\cite{Kim2015Deep} & 2015  &NITRC\tnote{4} & 50&50& rs-fMRI& DNN& 85.8\\\hline
			Patel~\etal~\cite{Patel2016Classification}& 2016 &COBRE\tnote{5}  &72&74&rs-fMRI &SAE &92.0 \\	\hline	
						Zeng~\etal~\cite{Zeng2018Multi}& 2018 &Multi-site2\tnote{6}&357  & 377&rs-fMRI &SAE &85.0$\pm$1.2\\\hline
              Qureshi ~\etal~\cite{qureshi20193d} & 2019 &COBRE &72&74&rs-fMRI &3D-CNN &98.09$\pm$1.01 \\	\hline	
			Dakka~\etal~\cite{dakka2017learning}& 2017&FBIRN\tnote{7}&46&49&rs-fMRI& CNN+LSTM&66.4\\\hline
			              Yan~\etal~\cite{yan2019discriminating}& 2019 &Multi-site3\tnote{8}&558&542&rs-fMRI &CNN+GRU&83.2$\pm$3.2 \\\hline
			Qi~\etal~\cite{Qi2016Deep}& 2016 &MLSP2014& 69&75&sMRI+fMRI &DCCA/DCCAE &94.2/95.0(AUC)\\\hline
			Srinivasagopalan~\etal~\cite{ srinivasagopalan2019deep}&2019&MLSP2014& 69&75&sMRI+fMRI&DNN&94.44\\\hline
			Ulloa~\etal~\cite{ulloa2018improving}&2018&FBIRN&135&169&sMRI+fMRI & DNN& 85.0$\pm$5.0(AUC)\\\hline

		\end{tabular*}
		\label{tab_NBH_Resluts}
		\begin{tablenotes}
			\item[1] Johns Hopkins University; the Maryland Psychiatric Research Center; the Institute of Psychiatry;  the Western Psychiatric Institute and Clinic at the University of Pittsburgh.
			\item[2] the Universidade Federal de São Paulo.
			\item[3] Northwestern University Schizophrenia Data and Software Tool.
			\item[4] Neuroimaging Informatics Tools and Resources Clearinghouse website.
			\item[5] Center for Biomedical Research Excellence.
			\item[6] Xijing Hospital; First Affliated Hospital of Anhui Medical University; Second Xiangya Hospital;  COBRE; the University of California, Los Angles and Washington University School of Medicine.
			\item[7] The Function Biomedical Informatics Research Network Data.
			\item[8] Peking University Sixth Hospital; Beijing Huilongguan Hospital; Xinxiang Hospital; Xinxiang Hospital; Xijing Hospital; Renmin Hospital of Wuhan University; Zhumadian Psychiatric Hospital.

		\end{tablenotes}
	\end{threeparttable}
	\label{tab:SZ01}
\end{table}

The fMRI-methods extracted discriminative features from rs-fMRI brain images with functional connectivity networks.
Kim~\etal~learned lower-to-higher features via the DNN model, of which each hidden layer was added $L_{1}$-regulation to control the weight sparsity, and achieve 85.8\% accuracy~\cite{Kim2015Deep}.
Patel~\etal~ used an SAE model with four hidden layers to separately train on each brain region. The input layer directly uses the complete time series of all active voxels without converting them into region-wise mean time series. Thus, this made that the model retained more information~\cite{Patel2016Classification}.
Due to the limited size of SZ dataset, Zeng~\etal~collected a large multi-site rs-fMRI dataset from seven neuroimaging resources. An SAE  with an optimized discriminant item was designed to learn imaging site-shared functional connectivity features. This model can achieve accurate SZ classification performance across multiple independent imaging sites, and the learned features found that dysfunctional integration of the cortical-striatal-cerebellar circuit may play an important role in SZ~\cite{Zeng2018Multi}.
Qureshi~\etal~built  a 3D-CNN based deep learning classification framework, which used the 3D ICA functional network maps as input. These ICA maps served as highly discriminative 3D imaging features for the discrimination of SZ~\cite{qureshi20193d}.
To exploit both spatial and temporal information, Dakka~\etal~ and Yan~\etal~proposed a recurrent convolutional neural network involving CNN followed by LSTM and GRU, respectively. The CNN extracted spatial features, which then were fed to the followed RNN model to learn the temporal dependencies~\cite{dakka2017learning,yan2019discriminating} .

As known to all, combined multi-modality brain images can improve the performance of disorder diagnosis. The MLSP2014 (Machine Learning for Signal Processing) SZ classification challenge provided 75 NC and 69 SZ which both contained sMRI and rs-fMRI brain images.
Qi~\etal~used deep canonical correlation analysis (DCCA) and deep canonically correlated auto-encoders (DCCAE) to fuse multi-modality features~\cite{Qi2016Deep}.
But in the proposed method, two modalities features directly were combined as 411 dimensional vector, then fed to 3-layer DNN model~\cite{ srinivasagopalan2019deep}.
To alleviate the missing modality, the synthetic sMRI and rs-fMRI images were generated by a generator proposed, and then were used to train a multi-modalities  DNN~\cite{ulloa2018improving}.
For clarity, the important information of the above-mentioned papers was summarized in Table \ref{tab:SZ01}.
From this table, it can be seen the datasets for SZ diagnosis come from different universities, hospitals and medical centers.

\section{Discussion and Future Direction}
\label{S4}
Although deep learning models had achieved great success in the field of neuroimaging-based brain disorder analysis, there are still some challenges that deserve further investigation. We summarize these potential challenges as follows and explore possible solutions.

Firstly, deep learning methods require a large number of samples to train neural networks, while it's usually difficult to acquire training samples in many real-world scenarios, especially for neuroimaging data. The lack of sufficient training data in neuroimage analysis has been repeatedly mentioned as a challenge to apply deep learning algorithms. To address this challenge, data augmentation strategy has been proposed and widely used to enlarge the number of training samples. In addition, the use of transfer learning~\cite{cheng2017multidomaion,cheng2015Multimodal} provides another solution, by transferring well-trained networks on big sample datasets (related to the to-be-analyzed disease) to a small sample dataset for further training.

Secondly, the missing data problem is unavoidable in multimodal neuroimaging studies, because subjects may lack some modalities due to patient dropouts and poor data quality. Conventional methods typically discard data-missing subjects, which will significantly reduce the number of training subjects and degrade the diagnosis performance. Although many data imputing methods have been proposed, most of them focus on imputing missing hand-crafted feature values that are defined by experts for representing neuroimages, while the hand-crafted features themselves could be not discriminative for disease diagnosis and prognosis. Several recent studies~\cite{Pan2018Synthesizing, pan2019disease} propose to directly impute missing neuroimages (\eg, PET) based on another modality neuroimages (\eg, MRI), while the correspondence between imaging data and non-imaging data has not been explored. We expect to see more deep network architectures in the near future to explore the association between different data modalities for imputing those missing data. 

Thirdly, an effective fusion of multimodal data has always been a challenge in the field. Multimodal data reflects the morphology, structure and physiological functions of normal tissues and organs from different aspects, and has strong complementary characteristics between different models. Previous studies for multimodal data fusion can be divided into two categories, \textit{data-level fusion} (focus on how to combine data from different modalities) and \textit{decision-level fusion} (focus on ensembling classifiers). Deep neural network architectures allow a third form of multimodal fusion, \ie, the intermediate fusion of learned representations, offering a truly flexible approach to multimodal fusion. As deep-learning architectures learn a hierarchical representation of underlying data across its hidden layers, learned representations between different modalities can be fused at various levels of abstraction. Further investigation is desired to study which layer of deep integration is optimal for problems at hand.

Furthermore, different imaging modalities usually reflect different temporal and spatial scales information of the brain. For example, sMRI data reflect minute-scale time scales information of the brain, while fMRI data can provide second-scale time scales information. In the practical diagnosis of brain disorder, it shows great significance for the implementation of early diagnosis and medical intervention by correctly introducing the spatial relationship of the diseased brain regions and other regions and the time relationship of the development of the disease progress~\cite{wang2019spatial,jie2018developing}. Although previous studies have begun to study the pathological mechanisms of brain diseases on a broad temporal and spatial scales, those methods usually consider either temporal or spatial characteristics. Therefore, it is desired to develop a series of deep learning frameworks to fuse temporal and spatial information for automated diagnosis of brain disorder.

Finally, the utilization of multi-site data for disease analysis has recently attracted increasing attention~\cite{Heinsfeld2018Identification,wang2018Low,wang2019identifying}, since a large number of subjects from multiple imaging sites are beneficial for investigating the pathological changes of disease-affected brains. Previous methods often suffer from inter-site heterogeneity caused by different scanning parameters and subject populations in different imaging sites, by assuming that these multi-site data are drawn from the same data distribution. Constructing accurate and robust learning models using heterogeneous multi-site data is still a challenging task. To alleviate the inter-site data heterogeneity, it could be a promising way to simultaneously learn adaptive classifiers and transferable features across multiple sites.


\section{Conclusion}

In this paper, we reviewed the most recent studies on the subject of applying the deep learning techniques in neuroimaging-based brain disorder analysis, and focused on four typical disorders. AD and PD are both neurodegenerative disorder. ASD and SZ are neurodevelopmental and psychiatric disorders, respectively. Deep learning models have achieved state-of-the-art performance across the four brain disorders using brain images.
Finally, we summarize these potential challenges and discuss possible research directions.
With the clearer pathogenesis of human brain disorders, the further development of deep learning techniques, the larger size of open source datasets, a human-machine collaboration for medical diagnosis and treatment will ultimately become a symbiosis in the future.

\section{Acknowledgments} This work was supported in part by National Natural Science Foundation of China (NSFC) under grants (Nos.~61802193, 61876082, 61861130366, and 61473149), the Natural Science Foundation of Jiangsu Province under grants (BK20170934), the Royal Society-Academy of Medical Sciences Newton Advanced Fellowship (No.~NAF$\backslash$R1$\backslash$180371), and the Fundamental Research Funds for the Central Universities (No.~NP2018104),

\footnotesize
\bibliographystyle{unsrt}
\bibliography{Review.bib}
\end{document}